\documentclass[aps,prb,twocolumn,amsmath,amssymb,nofootinbib,eqsecnum,superscriptaddress,floatfix]{revtex4}

\usepackage[dvips]{color} 
\usepackage{dcolumn} 
\usepackage{bm} 
\usepackage{amsmath}
\usepackage{amsthm}
\usepackage{amssymb}
\usepackage[mathcal]{euscript}
\usepackage{graphicx}

\begin{document}

\title{ 
Landauer conductance 
and twisted boundary conditions
for Dirac fermions 
in two space dimensions
      }

\author{
S.\ Ryu
       } 
\affiliation{
Kavli Institute for Theoretical Physics,
University of California, Santa Barbara, CA 93106, USA
            } 
\author{
C.\  Mudry
       } 
\affiliation{
Condensed matter theory group, 
Paul Scherrer Institute, 
CH-5232 Villigen PSI,
Switzerland
       }
\author{A.\ Furusaki
            } 
\affiliation{
Condensed Matter Theory Laboratory,
RIKEN, Wako, Saitama 351-0198, Japan
            }

\author{A.\ W.\ W.\ Ludwig} 
\affiliation{
Department of Physics, University of California, Santa Barbara, CA 93106, USA
            }
\affiliation{
Kavli Institute for Theoretical Physics, 
University of California, Santa Barbara, CA 93106, USA
            }

\date{\today}

\begin{abstract}
We apply the generating function technique developed by Nazarov
to the computation of the density of transmission eigenvalues 
for a two-dimensional free massless Dirac fermion,
which, e.g.,  underlies theoretical descriptions of graphene.
By modeling ideal leads attached to the sample as
a conformal invariant boundary condition, we relate
the generating function for the density of transmission eigenvalues 
to the twisted chiral partition functions
of fermionic ($c=1$) and bosonic ($c=-1$)
conformal field theories.
We also discuss the scaling behavior of the
ac Kubo conductivity and compare its \textit{different} dc limits
with results obtained from the Landauer conductance.
Finally, we show that the disorder averaged Einstein conductivity 
is an analytic function of the disorder strength, 
with vanishing first-order correction,
for a tight-binding model
on the honeycomb lattice with weak real-valued 
and nearest-neighbor random hopping.
\end{abstract}

\maketitle

\section{
Introduction
        }
\label{sec: intro}

The recent manufacture of a
single atomic layer of graphite (graphene)
has renewed interest in the transport properties 
of Dirac fermions propagating in two-dimensional space.
\cite{%
Novoselov04,%
Novoselov05,%
Zhang05a,%
Zhang05b%
     }
Recent theoretical work includes,
among many others, the computation of the
Landauer conductance for a single massless Dirac fermion
by Katsnelson in Ref.\ \onlinecite{Katsnelson06},
as well as the computation of the Landauer conductance 
and of  the Fano factor in Ref.~\onlinecite{Tworzydlo06},
confirming the result of
Ref.\ \onlinecite{Katsnelson06}
and predicting sub-Poissonian shot noise.

The conductivity has long been known to be related 
to a twist of boundary conditions.%
~\cite{Edwards71}
This idea has been further developed by Nazarov
who proposed a generating function for 
the density of transmission eigenvalues
in quasi-one-dimensional disordered conductors.%
~\cite{%
Nazarov94,%
Rejaei96,%
Brouwer96%
      }
With this formalism, Lamacraft, Simons, and Zirnbauer
reproduced in Ref.\ \onlinecite{Lamacraft04}     
(see also Ref.~\onlinecite{Altland05})
nonperturbative results
of
Refs.%
~\onlinecite{%
unitary Q1D wire,%
Mudry99,%
Brouwer00%
}
for the mean conductance and the density of transmission eigenvalues
of quasi-one-dimensional disordered quantum wires 
for three symmetry classes of Anderson localization.%

The purpose of this paper
is to establish a connection between (i)
the density of transmission eigenvalues 
of the noninteracting Dirac Hamiltonian describing the free (ballistic)
propagation of a relativistic massless electron in two-dimensional 
space and
(ii) twisted chiral partition functions of
a combination (tensor product)
of two conformal field theories (CFTs)
with central charges $c=1$ and $c=-1$.
In this way we provide a complementary method for calculating
the Landauer conductance of a single massless Dirac fermion
which agrees with the direct calculations
of Refs.~\onlinecite{Katsnelson06}
and~\onlinecite{Tworzydlo06},
while it might give us a powerful tool to account for the 
nonperturbative effects for certain types of disorder.
This connection comes from the observation that 
ideal leads attached to the sample, 
which are necessary to define the Landauer conductance,
can be replaced by a set of conformally invariant boundary conditions.
That this is possible is a consequence of the ``diffusive'' 
nature of the ballistic Dirac transport, 
as epitomized by its sub-Poissonian shot noise,
which in turn suggests an insensitivity to the modeling of ideal leads.

The transport properties of noninteracting Dirac fermions in
two-dimensional space under ideal conditions, i.e.,
without the breaking of translation invariance by disorder, 
have also been discussed in terms of the Kubo formula.
However, extracting the dc value of the conductivity
from the Kubo formula, which depends on
frequency $\omega$, temperature $1/\beta$, and 
smearing $\eta$ (imaginary part of the self-energy),
is rather subtle, as it is known that 
the conductivity in the dc limit
$(\omega, 1/\beta,\eta)=(0,0,0)$ is sensitive
to how one approaches the dc limit.%
\cite{Ludwig94,Fradkin86,Lee93,Shon98,Durst99,Gorbar02,
Peres05,Cserti06, Ziegler07}
There have been at least two known limiting procedures:
The Einstein conductivity, defined by 
first taking the zero temperature and 
then the zero frequency limits
while keeping $\eta$ finite,
is given by $1/\pi$ in units of $e^2/h$.%
\cite{Ludwig94,Fradkin86,Lee93,Shon98,Durst99,Gorbar02,
Peres05,Cserti06, Ziegler07}
On the other hand, by switching the order of zero smearing 
and zero frequency limits, 
one obtains, instead, $\pi/8$ 
for the dc conductivity (again in units of $e^2/h$).%
\cite{Ludwig94,Cserti06}

To clarify the origin of this sensitivity to how the dc limit
is taken, we compute the ac Kubo conductivity 
without taking any biased limit in $\omega$, $1/\beta$, or $\eta$.
Due to the scale invariance of noninteracting Dirac fermions, we show 
the Kubo conductivity to be a scaling function of two scaling variables.
We discuss several limiting procedures, including the above two,
and clarify the relationship between different dc values. 
In particular, we demonstrate that, if we take the zero smearing
limit prior to the other two limits, 
we can obtain any value between 0 and $\pi/8$ 
for the dc Kubo conductivity.
The Einstein conductivity agrees with the conductivity determined 
from the Landauer conductance. 
We also compute several asymptotic behaviors of the Kubo conductivity 
that have not been obtained before. 
These considerations may be of relevance to experiments 
on graphene if different limiting procedures are accessed.

The perturbative effects of disorder in the form of weak real-valued random
(white-noise) hopping between nearest-neighbor sites of the honeycomb lattice
at the band center are discussed. 
We show that, as a consequence of the fixed point
theory discussed in Ref.~\onlinecite{Guruswamy00},
the Einstein conductivity is an analytic function of the disorder strength.
We also show that the first-order 
correction to the Einstein conductivity vanishes, 
in agreement with a calculation performed by 
Ostrovsky et al.\ in Ref.~\onlinecite{Ostrovsky06}.
This result is of relevance to 
the two-dimensional chiral-orthogonal universality class.%
\cite{Verbaarschot94,Zirnbauer96,Altland97,Heinzner05} 
Potential relationships of the two-dimensional chiral symmetry 
class with certain types of disorder in graphene
were recently discussed in Ref.\  \onlinecite{Ostrovsky06}.
We refer the reader to 
Refs.~\onlinecite{Ostrovsky06}, \onlinecite{Aleiner06}, and \onlinecite{Altland06}
for a discussion of white-noise disorder in graphene 
in terms of symmetry classes and to 
Refs.~\onlinecite{Aleiner06,Altland06,Ando02,Rycerz07,Nomura07}
for the possibility that smooth disorder could induce 
crossovers between different symmetry classes.

\section{
Model
        }
\label{sec: Definitions and results}

Our starting point is the single-species (or one-flavor) Dirac Hamiltonian
\begin{equation}
\begin{split}
\mathcal{H}:=
-{i}
\hbar v^{\ }_F\,
\sigma^{\ }_{\mu}\,
\partial^{\ }_{\mu},
\qquad
\hat H :=
\int d^2 r\,
\hat\Psi^{\dag}
\mathcal{H}
\hat\Psi^{\ },
\end{split}
\label{eq: def Dirac H}
\end{equation}
where 
$\hat\Psi^{\dag}$ ($\hat\Psi$) is a two-component fermionic creation (annihilation) operator
and $v^{\ }_F$ the Fermi velocity.
We choose $\sigma^{\ }_{\mu=x,y}$ to be the first two of the three Pauli
matrices
$\sigma^{\ }_{x}$, $\sigma^{\ }_{y}$, and $\sigma^{\ }_{z}$ in the
standard representation.
(We use the summation convention over repeated indices.)
Hamiltonian~(\ref{eq: def Dirac H}) describes the free relativistic
propagation of a spinless fermion in two-dimensional space parametrized by
the coordinates $r=(x,y)$. 
As such, it possesses the chiral symmetry 
\begin{equation}
\sigma^{\ }_{z}\,
\mathcal{H}\,
\sigma^{\ }_{z}=
-\mathcal{H}.
\label{eq: chiral sym}
\end{equation}

The single-particle retarded and advanced Green's functions are defined by
\begin{equation}
\mathcal{G}^{{R}/{A}}_{\eta}(\varepsilon):=
(\varepsilon \pm{i}\hbar\eta-\mathcal{H})^{-1}.
\label{eq: def retarded Green fct}
\end{equation}
Consequently, at the band center $\varepsilon=0$, 
they are related to each other by the chiral transformation as
\begin{equation}
\sigma^{\ }_z\, 
\mathcal{G}^{{R}}_{\eta}(\varepsilon=0)\, 
\sigma^{\ }_z=
-
\mathcal{G}^{{A}}_{\eta}(\varepsilon=0).
\label{eq: chiral symmetry for green function}
\end{equation}
Below, matrix elements between eigenstates of the position operator
of the single-particle retarded Green's function evaluated at
$\varepsilon$ are denoted by $\mathcal{G}^{{R}}_{\eta}(r,r';\varepsilon)$.

In the presence of an electromagnetic vector potential $A^{\ }_{\mu}(r)$, 
one modifies Hamiltonian%
~(\ref{eq: def Dirac H})
through the minimal coupling
$
\partial^{\ }_{\mu}\to
\partial^{\ }_{\mu}-{i}(e/\hbar c)A^{\ }_{\mu}
$ (with $e<0$).
The conserved charge current then follows from taking the 
functional derivative with respect to $A^{\ }_{\mu}(r)$,
\begin{equation}
\begin{split}
\hat{j}^{\ }_{\mu}(r)=
\left(
\hat\Psi^{\dag}
j^{\ }_{\mu}
\hat\Psi
\right)(r),
\qquad
j^{\ }_{\mu} &:= 
ev^{\ }_{F}\,\sigma^{\ }_{\mu}.
\end{split}
\label{DefJmu}
\end{equation}
There is no diamagnetic contribution
due to the linear dispersion.

\section{
Kubo and Einstein conductivities
        }
\label{sec: Kubo and Einstein conductivities}

\subsection{
Linear response
           }

We start from the bilocal conductivity tensor 
at a finite temperature $1/\beta$
defined by the linear response
relation in the frequency-$\omega$ domain
\begin{subequations}
\label{eq: def bilocal conductivity}
\begin{equation}
\begin{split}
j^{\mathrm{ind}}_{\mu}(r,\omega,\beta)
=&
\int d^2 r'\,
\sigma^{\ }_{\mu\nu}(r,r';\omega,\delta,\beta)
E^{\ }_{\nu}(r',\omega)
\end{split}
\end{equation}
between the $\omega$ component 
$
E^{\ }_{\nu}(r',\omega)
$
of an electric field that has been switched on adiabatically
at $t=-\infty$ and the induced local current 
$j^{\mathrm{ind}}_{\mu}(r,\omega,\beta)$,
where~\cite{Mahan}
\begin{equation}
\sigma^{\ }_{\mu\nu}(r,r';\omega,\delta,\beta)=
\frac{
D^{{R}}_{\mu\nu}(r,r';\omega,\delta,\beta)
     }
     {
\hbar\omega
     }.
\label{eq: bilocal conductivity at omega}
\end{equation}
Here, 
\begin{equation}
  D^{{R}}_{\mu\nu}(r,r';\omega,\delta,\beta):=
  \int\limits_0^{\infty}\!
  dt\,
  e^{\mathrm{i}(\omega +{i}\delta)t}
  \left\langle
  \big[
  \hat{j}^{\ }_{\mu}(r,t),\hat{j}^{\ }_{\nu}(r')
  \big]
  \right\rangle^{\ }_{\beta}
\label{eq: responce function}
\end{equation}
\end{subequations}
is the response function,
$\hat{j}_{\mu}(r,t)$ is the current operator
in the Heisenberg picture,
and $\langle \cdots \rangle^{\ }_{\beta}$
is the expectation value taken with respect to 
the equilibrium density matrix at temperature $1/\beta$.
The small positive number $\delta >0$
implements the adiabatic switch-on of the electric field.
The conductivity tensor in a sample of linear size $L$ is defined by 
integrating over the spatial coordinates of 
the bilocal conductivity tensor,
\begin{equation}
\sigma^{\ }_{\mu\nu}(\omega,\delta,\beta,L):=
\int \frac{d^{2}r}{L^{2}}
\int d^{2}r'
\sigma^{\ }_{\mu\nu}(r,r';\omega,\delta,\beta).
\label{eq: def conductivity tensor if single particle H}
\end{equation}

We impose periodic boundary conditions
and choose to represent
the Dirac Hamiltonian%
~(\ref{eq: def Dirac H})
by
\begin{subequations}
\begin{eqnarray}
\hat{H}= 
\sum_{p}
\sum_{\sigma=\pm}
\sigma 
\varepsilon^{\ }_{p} 
\hat a^{\dag} _{p,\sigma}
\hat a^{\ }_{p,\sigma},
\end{eqnarray}
where the fermionic creation operators
$\hat a^{\dag}_{p,\sigma}$
with $\sigma=\pm$
create from the Fock vacuum $|0\rangle$
the single-particle eigenstates 
with momentum $p^{\ }_{\mu}$
\begin{equation}
|m\rangle\equiv
|p,\sigma\rangle\equiv
\hat{a}^{\dag}_{p,\sigma}|0\rangle:=
\frac{1}{\sqrt{2}|p|}
\left(
\begin{array}{c}
\sigma
\left(
p^{\ }_{x}-{i}p^{\ }_{y}
\right)
\\
|p|
\end{array}
\right),
\label{eq: single-particle eigenstates}
\end{equation}
of $\mathcal{H}$ 
with the single-particle energy eigenvalues $\varepsilon^{\ }_{m}$,
\begin{equation}
\varepsilon^{\ }_{m}\equiv
\sigma \varepsilon^{\ }_{p}=
\sigma
v^{\ }_{F}|p|\equiv
\sigma
v^{\ }_{F}\sqrt{p^{2}_{x}+p^{2}_{y}}.
\end{equation}
\end{subequations}
The conductivity tensor 
can then be expressed solely in terms
of single-particle plane waves
\begin{equation}
\begin{split}
\sigma^{\ }_{\mu\nu}(\omega,\delta,\beta,L)=&\,
\frac{{i}}{\omega}
\int d\varepsilon\,
\int \frac{d^2 r}{L^2}
\int d^2 r'
\\
&
\times
\sum_{m,n}
\langle m|\hat{j}^{\ }_{\mu}(r )|n\rangle
\langle n|\hat{j}^{\ }_{\nu}(r')|m\rangle
\\
&
\times
\delta(\varepsilon-\varepsilon^{\ }_{m})
\frac{
f^{\ }_{\beta}(\varepsilon)
-
f^{\ }_{\beta}(\varepsilon^{\ }_{n})
     }
     {
\varepsilon
-
\varepsilon^{\ }_{n}
+
\hbar
\left(
\omega
+
{i}\delta
\right)
     },
\label{eq: bilocal if single particle H}
\end{split}
\end{equation}
where 
$f^{\ }_{\beta}(\varepsilon):=1/(e^{\beta\varepsilon}+1)$
is the Fermi-Dirac function at temperature $1/\beta$
and at zero chemical potential.

As usual, the infinite-volume limit
$L\to \infty$ has to be taken before the
$\delta\to 0$ limit in Eq.\ (\ref{eq: bilocal if single particle H}).
(Recall that $\delta$ controls the adiabatic switching of the external field.)
In the following, it is understood that
we always take these limits prior to any other limits.
We thus drop the explicit $\delta$ and $L$ dependence
of the conductivity tensor henceforth.
The dc conductivity can then be computed by
taking the subsequent limit, $\omega \to 0$.
The temperature $1/\beta$ can be fixed to some arbitrary value.

The real part of Eq.\ (\ref{eq: bilocal if single particle H})
can be further rewritten in terms of single-particle Green's
functions. 
This can be done by first replacing the two $\delta$ functions in 
the real part of
Eq.\ (\ref{eq: bilocal if single particle H}),
which appear after taking the $\delta\to0$ limit, 
by two Lorentzians with the same width $\hbar\eta$
(see, for example, Ref.~\onlinecite{Baranger89}).
Then, each Lorentzian can be rewritten as
the difference of the retarded and advanced Green's functions,
$\mathcal{G}^{R}_{\eta}(\varepsilon)-\mathcal{G}^{A}_{\eta}(\varepsilon)$.
By also noting that
the transverse components ($\mu\neq \nu$)
of the conductivity tensor (\ref{eq: bilocal if single particle H})
vanish by the spatial symmetries of the matrix elements in
$
\int d^2r \int d^2r'
\langle m|\hat{j}^{\ }_{\mu}(r )|n\rangle
\langle n|\hat{j}^{\ }_{\nu}(r')|m\rangle
$,
we obtain
\begin{equation}
\begin{split}
\mathrm{Re}\,
\sigma^{\ }_{\mu\nu}(\omega,\eta,\beta)=&\,
\delta^{\ }_{\mu\nu}\,
\frac{\hbar}{4\pi} 
\int d\varepsilon
\frac{ 
f^{\ }_{\beta}(\varepsilon+\hbar\omega)
-
f_{\beta}(\varepsilon)
     }
     {
\hbar\omega
     }
\\
&
\times
\int \frac{d^2 r}{L^2}
\int d^2 r'
\Sigma^{\ }_{\mu\nu}
(r,r';\varepsilon,\omega,\eta),
\end{split}
\label{eq: bilocal if single particle H 2}
\end{equation}
where we have introduced
\begin{equation}
\begin{split}
\Sigma^{\ }_{\mu\nu}
(r,r';\varepsilon,\omega,\eta):=&
\,
\mathrm{tr}\,
\Big[
\mathcal{G}^{{A}-{R}}_{\eta}(r',r;\varepsilon) 
j^{\ }_{\mu}
\\
&
\times
\mathcal{G}^{{A}-{R}}_{\eta}(r,r';\varepsilon+\hbar\omega) 
j^{\ }_{\nu}
\Big]
\label{eq: bilocal conductivity at omega 0 in terms Greens functions}
\end{split}
\end{equation}
and
$
j^{\ }_{\mu}
$
is defined in Eqs.\ (\ref{DefJmu}).
Here
the
trace is 
taken over spinor indices and
\begin{equation}
\mathcal{G}^{{A}-{R}}_{\eta}(r,r';\varepsilon):= 
\mathcal{G}^{{A}}_{\eta}(r,r';\varepsilon)
-
\mathcal{G}^{{R}}_{\eta}(r,r';\varepsilon).\
\label{GreensAR}
\end{equation}
Using translational invariance,
the single-particle Green's functions are given
by\cite{footnote: a dot b}
\begin{equation}
\begin{split}
  \mathcal{G}^{R/A}_{\eta}(r^{\ }_1,r^{\ }_2;\varepsilon)
  =&
  \int_{k}
  e^{+{i}k\cdot (r^{\ }_{2}-r^{\ }_1)}
  \mathcal{G}^{R/A}_{\eta} (k;\varepsilon),
\\
  \mathcal{G}^{R/A}_{\eta}(k;\varepsilon)
  =&
  \frac{1}{\varepsilon\pm {i}\hbar \eta -\hbar v^{\ }_F k \cdot \sigma}.
\end{split}
\label{eq: Green function}
\end{equation}

In order to define the conductivity in the clean system
we should take the $\eta\to 0$ limit
before the $\omega\to 0$ limit.
On the other hand, $\eta$ can be interpreted
physically as a finite inverse life time 
(imaginary part of the self energy)
induced by disorder.
Thus, it is meaningful to discuss
Eq.\ (\ref{eq: bilocal if single particle H 2}) 
in the presence of finite $\eta$.
Below, we first discuss 
the $\eta\to 0$ limit.
We will then discuss the case of 
finite $\eta$.

\subsection{$\eta\to 0$ limit}

We define the ac Kubo conductivity tensor 
at any finite frequency $\omega >0$
and temperature $1/\beta$ by
\begin{equation}
\begin{split}
\sigma^{\mathrm{K}}_{\mu\nu}(\omega,\beta):=&
\lim_{\eta\to0}
\mathrm{Re}\,
\sigma^{\ }_{\mu\nu}(\omega,\eta,\beta).
\end{split}
\label{eq: def ac Kubo conductivity}
\end{equation}
With the help of Eq.%
~(\ref{eq: bilocal if single particle H}),
\begin{equation}
\begin{split}
\sigma^{\mathrm{K}}_{\mu\nu}(\omega,\beta)=&
  \frac{\pi(ev^{\ }_{F})^{2}}{\omega}
  \sum_{m,n}
  \langle m|\sigma^{\ }_{\mu}|n\rangle
  \langle n|\sigma^{\ }_{\nu}|m\rangle
\\
&
\times
  \left[
    f^{\ }_{\beta}(\varepsilon^{\ }_{m})
    -
    f^{\ }_{\beta}(\varepsilon^{\ }_{n})
  \right]
\delta
\left(
\varepsilon^{\ }_{m}-\varepsilon^{\ }_{n}+\hbar\omega
\right).
\label{eq: ac Kubo conductivity a}
\end{split}
\end{equation}

When $\omega > 0$ and $\beta < \infty$,
the sum over the basis%
~(\ref{eq: single-particle eigenstates})
in the real part of Eq.~(\ref{eq: ac Kubo conductivity a})
can be performed once the matrix elements of the currents
have been evaluated, yielding
\begin{eqnarray}
\sigma^{\mathrm{K}}_{\mu\nu}(\omega,\beta)&=&
\delta^{\ }_{\mu\nu}
\frac{e^{2}}{h}\frac{\pi}{8}
\tanh\frac{\beta\hbar \omega}{4}.
\label{eq: ac Kubo conductivity b}
\end{eqnarray}
Observe that Eq.~(\ref{eq: ac Kubo conductivity b})
is independent of the Fermi velocity $v^{\ }_{F}$.%
\cite{footnote: role of zero modes}
Finally, the ac Kubo conductivity%
~(\ref{eq: ac Kubo conductivity b}) 
depends solely on the combination
\begin{equation}
Z:=
\beta\hbar \omega \in\mathbb{R}.
\label{eq: def scaling variable x}
\end{equation}

The limiting value of Eq.~(\ref{eq: ac Kubo conductivity b})
when $\omega\to0$ and $\beta\to\infty$ can be any number between
0 and $e^{2}\pi/(8h)$
provided the scaling variable~(\ref{eq: def scaling variable x})
is held fixed.
For example,
if the limit $\beta\to\infty$ is taken before the limit $\omega\to0$,
then
\begin{equation}
\lim_{\omega\to0}
\lim_{\beta\to \infty}
\sigma^{\mathrm{K}}_{\mu\nu}(\omega,\beta)=
\delta^{\ }_{\mu\nu}
\frac{e^2}{h}\frac{\pi}{8}.
\label{eq: order of limit I for dc Kubo}
\end{equation}
This limiting procedure reproduces the results from Refs.%
~\onlinecite{Ludwig94}
and~\onlinecite{Cserti06}.
On the other hand, if the limit $\omega\to 0$
is taken before the limit $\beta\to0$, then
\begin{equation}
\lim_{\beta\to\infty}
\lim_{\omega\to 0}
\sigma^{\mathrm{K}}_{\mu\nu}(\omega,\beta)=
0.
\label{eq: order of limit II for dc Kubo}
\end{equation}
Clearly,
$
\lim_{\beta \to \infty}\sigma^{\mathrm{K}}_{\mu\nu}(\omega,\beta)=
\delta^{\ }_{\mu\nu}
(\pi/8) (e^2/h) 
$
for any finite frequency $\omega$, while
$
\lim_{\omega\to 0}\sigma^{\mathrm{K}}_{\mu\nu}(\omega,\beta)=
0
$
for any finite temperature $1/\beta$.
The singularity at $\omega=1/\beta=0$ is a manifestation of
the linear dispersion of the massless Dirac spectrum
leading to a dependence on 
the scaled variables $\omega$ and $\beta$.

\subsection{Case of finite $\eta >0$}

For $\eta$ finite,
it is shown in Appendix~\ref{app sec: preparation for numerical integration}
that the real part of the longitudinal conductivity, 
Eq.\ (\ref{eq: bilocal if single particle H 2}), 
is a scaling function of two variables, i.e.,
\begin{equation}
\mathrm{Re}\,\sigma^{\ }_{xx}(\omega,\eta,\beta)=
\mathrm{Re}\,\sigma^{\ }_{xx}(X,Y),
\end{equation}
where
\begin{eqnarray}
X:=\frac{\omega}{\eta},
\qquad
Y:=\frac{1}{\beta\hbar\eta}.
\label{eq: def scaling variables x and y}
\end{eqnarray}

\subsubsection{
dc response $\omega=0$
              }

If we take the dc limit $\omega\to0$ while keeping $\eta>0$ finite,
Eq.\ (\ref{eq: bilocal if single particle H 2}) 
can be expressed as
\begin{equation}
\begin{split}
\mathrm{Re}\,\sigma^{\ }_{\mu\nu}(X=0,Y)=&\,
\delta^{\ }_{\mu\nu}
\frac{\hbar}{4\pi} 
\int d \varepsilon
\frac{\partial f^{\ }_{\beta}}{\partial\varepsilon}
\int \frac{d^{2}r}{L^{2}}
\int d^{2}r'
\\
&
\times
\Sigma^{\ }_{\mu\nu}
(r,r';\varepsilon,\omega=0,\eta),
\end{split}
\label{eq: basis independent representation Einstein conductivity}
\end{equation}
where
$\Sigma^{\ }_{\mu\nu}$
is defined in Eqs.%
~(\ref{eq: bilocal conductivity at omega 0 in terms Greens functions})
and%
~(\ref{GreensAR}).
With the help of
[see Eqs.~(\ref{eq: def Dirac H}) and (\ref{DefJmu})]
\begin{equation}
j^{\ }_{\mu}=
{i}e
\left[\mathcal{H},r^{\ }_{\mu}\right]/\hbar,
\label{eq: jmu is a commutator}
\end{equation} 
one can show that%
~\cite{Kramer93}
\begin{equation}
\begin{split}
\mathrm{Re}\,\sigma^{\ }_{\mu\nu}(X=0,Y)=&\,
\delta^{\ }_{\mu\nu}
  \frac{e^2}{h}
  \eta^2
\int d\varepsilon
\frac{\partial f_{\beta}}{\partial \varepsilon}
  \int
  d^2 r\,
  r^2\,
\\
&
\times
\mathrm{tr}\,
 \left[
  \mathcal{G}^{R}_{\eta} (0,r;\varepsilon)
  \mathcal{G}^{A}_{\eta} (r,0;\varepsilon)
 \right],
\end{split}
\label{eq: equivalence with Einstein cond at T=0}
\end{equation}
with $\mu=x,y$.
Equation (\ref{eq: equivalence with Einstein cond at T=0})
is usually referred to 
as the Einstein conductivity since 
it is related to the diffusion constant
via the Einstein relation
(see, for example, Ref.\ \onlinecite{McKane81}).

A closed-form expression for 
Eq.~(\ref{eq: basis independent representation Einstein conductivity})
can be obtained at zero temperature,
\begin{equation}
\mathrm{Re}\,\sigma^{\ }_{\mu\nu}(X=0,Y=0)=
\delta^{\ }_{\mu\nu} 
\frac{e^2}{h}
\frac{1}{\pi}.
\label{eq: order of limit III for dc Kubo}
\end{equation}
The same value was derived
in Ref.~\onlinecite{Ludwig94}.
Related predictions were also made, 
among others, in Refs.\
\onlinecite{Fradkin86},
\onlinecite{Lee93},
\onlinecite{Shon98},
\onlinecite{Durst99},
\onlinecite{Gorbar02},
\onlinecite{Peres05},
and
\onlinecite{Cserti06}.%
~\cite{footnote: Durst}
Equation (\ref{eq: order of limit III for dc Kubo})
also agrees with
the conductivity determined from the Landauer formula.
(This was first observed  in Ref.~\onlinecite{Katsnelson06}. 
We will reproduce this fact in
Sec.~\ref{sec: The Landauer conductance}
from  Nazarov's generating function technique.)

Whenever $\eta >0$,
Eq.\ (\ref{eq: bilocal if single particle H 2}) 
is an analytic function of $X$ and $Y$
at $(X,Y)=(0,0)$.
For $X,Y\ll 1$, 
the power series expansion of
Eq.\ (\ref{eq: bilocal if single particle H 2}) 
in terms of $X$ and $Y$
is given by
\begin{equation}
\begin{split}
\mathrm{Re}\,\sigma^{\ }_{xx}(X \ll 1,Y \ll 1)=
\frac{e^2}{h}
\frac{1}{\pi}
\left[
1
+
\left(\frac{X}{3}\right)^{2}
+
\left(\frac{\pi Y}{3}\right)^{2}
\right]
\end{split}
\label{eq: Kubo for small X and Y}
\end{equation}
up to terms of order $X^4$, $Y^4$,  or $X^{2}Y^{2}$.

\subsubsection{
Arbitrary $\omega$ and $\beta$ at finite $\eta$
           }

\begin{figure}
\begin{center}
\includegraphics[width=7.5cm,clip]{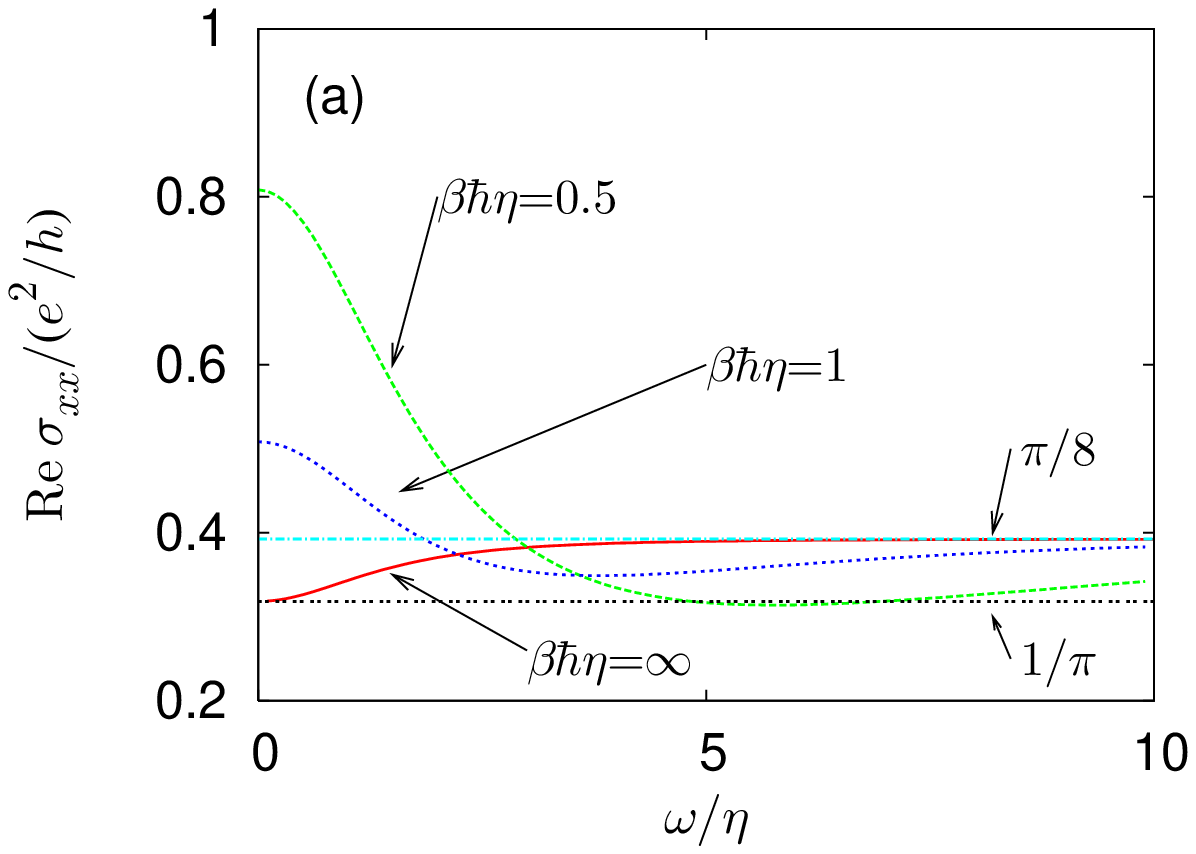}\\
\includegraphics[width=7.5cm,clip]{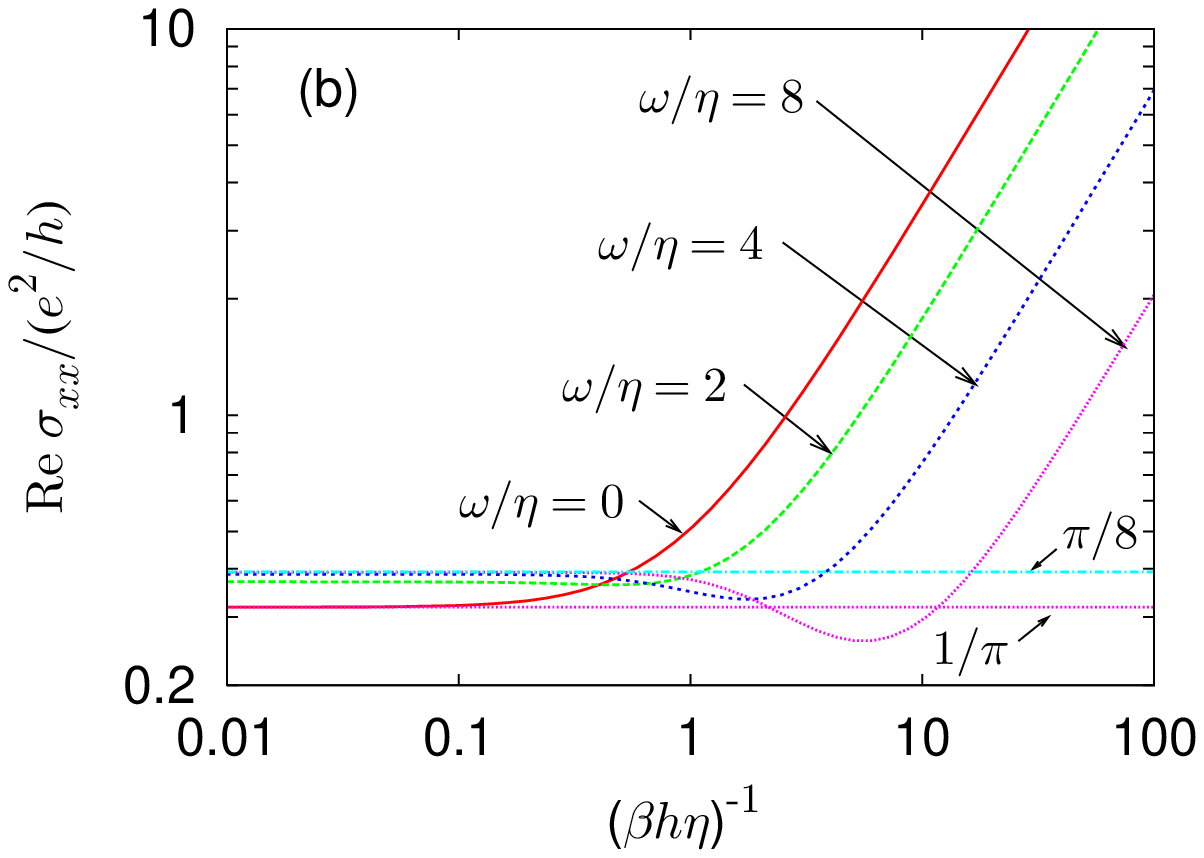}
\end{center}
\caption{
\label{fig: numerical integration}
(Color online)
Numerical integration of Eq.\ 
(\ref{eq: bilocal if single particle H 2}) 
when $\mu=\nu=x$
as a function of $X=\omega/\eta$ (a) 
and $Y=(\beta\hbar \eta)^{-1}$ (b).
The dc limit $\omega/\eta=0$ in (b)
is obtained from numerical integration of 
Eq.\ (\ref{eq: basis independent representation Einstein conductivity}).
        }
\end{figure}

For generic values of 
$(X=\omega/\eta,Y=1/\beta\hbar\eta)$,
we are unable to evaluate 
the real part of the longitudinal conductivity from 
Eq.\ (\ref{eq: bilocal if single particle H 2})
in closed form.
The numerical integration 
of Eq.\ (\ref{eq: bilocal if single particle H 2})
when $\mu=\nu=x$ or, equivalently,
of Eq.\ (\ref{eq: integrating over momenta a})
is presented in Fig.\ \ref{fig: numerical integration}.%
~\cite{OtherNumericsFigOne}

First, we fix $Y=1/(\beta\hbar\eta)$ 
and discuss the $X=\omega/\eta$ dependence of 
the real part of the longitudinal conductivity from 
Eq.\ (\ref{eq: bilocal if single particle H 2}).
We distinguish two limits.
When $X\to\infty$
as happens when the frequency $\omega$ is much larger than
the inverse life-time $\eta$, 
the real part of the longitudinal conductivity
measured in units of $e^{2}/h$ converges to the limiting
value $\pi/8$.
In the opposite limit of $X\to0$,
the limiting value is an increasing function of $Y$
and is given by
Eq.\ (\ref{eq: Kubo for small X and Y}) with $X=0$,
when the temperature $1/\beta$ is much smaller than the energy
smearing $\hbar\eta$ ($Y\ll1$).
These two limiting behaviors are smoothly connected as 
is illustrated in Fig.\ \ref{fig: numerical integration}(a).
For example,
$(h/e^2)\,\mathrm{Re}\,\sigma^{\ }_{xx}(X,Y=0)$
increases monotonically between
$1/\pi$ and $\pi/8$ 
as a function of $X$.\cite{Ostrovsky06}
The approach to the limiting value
$\pi/8$
for large $X$ when $Y=0$ is given by
\begin{equation}
\mathrm{Re}\,\sigma^{\ }_{xx}(X\gg1,Y=0)=
\frac{e^2}{h}
\left(
\frac{\pi}{8}
-
\frac{1}{3 X^3}
\right)
\end{equation}
up to terms of order $X^{-4}$.
For $Y\gg1$ the approach to the limit $X\to 0$ is Drude-like,
\begin{eqnarray}
\mathrm{Re}\,\sigma^{\ }_{xx}(X \ll 1 ,Y\gg 1)
\sim 
\frac{e^2}{h}
\frac{2Y \ln 2}{X^2 + 4}.
\end{eqnarray}

Second, we fix $X=\omega/\eta$
and discuss the $Y=1/(\beta\hbar\eta)$ dependence of 
the real part of the longitudinal conductivity from 
Eq.\ (\ref{eq: bilocal if single particle H 2}).
When $Y\gg1$ 
as happens when the temperature $\beta^{-1}$ 
is much larger than the energy resolution $\hbar\eta$, 
$
(h/e^2)\,\mathrm{Re}\,\sigma^{\ }_{xx}(X,Y)
\sim
gY
$
for any fixed value of $X$
where $g$ is some constant.
In particular, when $X=0$, we find
\begin{equation}
\mathrm{Re}\,\sigma^{\ }_{xx}(X=0,Y \gg 1)=
\frac{e^2}{h}\frac{\ln2}{2}\, Y
\end{equation}
to leading order in $Y\gg1$.\cite{Wallace47,Gorbar02}
In the opposite limit of $Y\ll1$,
the conductivity approaches a finite value
given by
Eq.\ (\ref{eq: Kubo for small X and Y})
with $Y=0$ when $X\ll1$,
which is thus an increasing function of $X\ll1$ in agreement 
with Fig.\ \ref{fig: numerical integration}(b).
The dependence on $Y$ of
$\mathrm{Re}\,\sigma^{\ }_{xx}(X,Y)$
is monotonic increasing if $X=0$
while it is nonmonotonic for the finite values of $X$
given in Fig.\ \ref{fig: numerical integration}(b).

\section{
Landauer conductance
           }
\label{sec: The Landauer conductance}

In this section we are going to reproduce the
calculation of the Landauer conductance for a
single massless Dirac fermion from
Refs.~\onlinecite{Katsnelson06}
and~\onlinecite{Tworzydlo06}
using the tools of CFT
subjected to boundary conditions that preserve 
conformal invariance.
Although the direct methods of
Refs.~\onlinecite{Katsnelson06}
and~\onlinecite{Tworzydlo06}
are both elegant and physically intuitive in the ballistic regime,
we are hoping that the CFT approach might lend itself
to a nonperturbative treatment of certain types of disorder.

\subsection{Definition}

In order to define the Landauer conductance,
we consider a finite region, the sample, 
described by the Dirac Hamiltonian,
and attach a set of leads (or reservoirs) 
$l^{\ }_1,l^{\ }_2,\cdots$ to the sample. 
Propagation in the leads obeys different laws than in the sample.%
~\cite{footnote: ideal leads}
Then, the (dimensionfull) 
conductance $G^{\mathrm{L}}_{a\to b}$ 
for the transport from the $a$th to $b$th lead
is determined from the transmission matrix $T^{\ }_{a\to b}$ by
\begin{equation}
G^{\mathrm{L}}_{a\to b}:=
\frac{e^{2}} {h}
\mathrm{tr}'
\left(T^{\dag}_{a \to b}T^{\ }_{a \to b}\right)
\label{eq: def Landauer cond}
\end{equation}
where $\mathrm{tr}'$ denotes
the trace over all channels 
in the $b$th lead. 
The Landauer conductance%
~(\ref{eq: def Landauer cond})
can be expressed in terms of the 
bilocal conductivity tensor 
$\Sigma^{\ }_{\mu\nu}$
of
Eq.\ (\ref{eq: bilocal if single particle H 2}),
according to%
~\cite{Fisher81,Baranger89,Xiong96} 
\begin{equation}
  G^{\mathrm{L}}_{a\to b}=
  -
  \int_{a} \int_{b}
  dS^{\mu}_{a}
  dS^{\prime \nu}_{b}\,
  \Sigma^{\ }_{\mu\nu}(r,r';\varepsilon=0,\omega=0,\eta=0),
  \label{eq: master formula for conductance in open geometry}
\end{equation}
where $r$ ($r'$) is constrained to lie on the interface between 
the $a$th ($b$th) lead and the sample,
and $\int_{a} dS^{\mu}_{a}$ represents integration over 
the oriented interface between the sample and the $a$th lead.

To define the longitudinal Landauer conductance 
we choose for the sample the surface of a cylinder of length $L^{\ }_{x}$ and of
perimeter $L^{\ }_{y}$ to which we attach two ideal leads $l^{\ }_1$ and $l^{\ }_2$
at the left end $x^{\ }_{L}=-L^{\ }_{x}/2$ and right end $x^{\ }_{R}=+L^{\ }_{x}/2$,
respectively.\cite{footnote: ideal leads}

For the free Dirac Hamiltonian
(\ref{eq: def Dirac H}),
the dimensionless conductance along the $x$ direction,
\begin{subequations}
\label{eq: Landauer conductance in 2 terminal geometry}
\begin{equation}
g^{\mathrm{L}}_{xx}=
(h/e^2)
G^{\mathrm{L}}_{xx}
\equiv 
(h/e^2)
G^{\mathrm{L}}_{1\to 2} ,
\label{eq: dimensionless conductance}
\end{equation}
can then be expressed 
in terms of the single-particle Green's function 
of Eq.\ (\ref{eq: def retarded Green fct})
as
\begin{equation}
\begin{split}
g^{\mathrm{L}}_{xx}=&
(\hbar v^{\ }_F)^2
\oint\limits^{0}_{L_y}dy
\oint\limits^{L^{\ }_{y}}_{0}dy'\,
\\
&
\times
\mathrm{tr} \left[
\mathcal{G}^{R}_{\eta=0}(r,r';0) \sigma^{\ }_{+} 
\mathcal{G}^{R}_{\eta=0}(r',r;0) \sigma^{\ }_{-}
\right],
\end{split}
\label{eq: longitudinal conductance}
\end{equation}
\end{subequations}
where $r=(x^{\ }_{L},y)$, $r'=(x^{\ }_{R},y')$, and
$\sigma^{\ }_{\pm}=\sigma^{\ }_{x}\pm {i}\sigma^{\ }_{y}$.
We made use of the chiral symmetry
and of 
$\Sigma^{\ }_{\mu\mu}(r,r';\varepsilon=0,\omega=0,\eta=0)
=\Sigma^{\ }_{\mu\mu}(r',r;\varepsilon=0,\omega=0,\eta=0)$.
The single-particle Green's functions that enter 
Eq.\ (\ref{eq: master formula for conductance in open geometry})
are obtained by solving the Schr\"odinger
equation for the entire system, including the leads.%
~\cite{footnote: ideal leads}  
For convenience, we assume that the leads also respect
chiral\cite{Verbaarschot94,Zirnbauer96,Altland97,Heinzner05} symmetry.
The imaginary part $\hbar\eta$ of the energy can be set to zero in the
sample, since the ideal leads broaden the energy levels in
the sample.  

In the sequel, we will use Nazarov's technique
to derive the following expressions for the
dimensionless conductance along the $x$ direction,
\begin{equation}
g^{\mathrm{L}}_{xx}=
\frac{1}{\pi}\frac{L^{\ }_{y}}{L^{\ }_{x}}
+
\mathcal{O}[(L^{\ }_{y}/L^{\ }_{x})^{0}].
\label{eq: result for gLxx}
\end{equation}
Thus, since the longitudinal conductivity
$\sigma^{\ }_{xx}$ can be extracted from the
conductance in the anisotropic limit via
$g^{\mathrm{L}}_{xx}=\sigma^{\ }_{xx} L^{\ }_{y}/L^{\ }_{x}$
where $L^{\ }_{x}\ll L^{\ }_{y}$, we recover
from (\ref{eq: result for gLxx})
the result
(\ref{eq: order of limit III for dc Kubo})
for the longitudinal Kubo dc conductivity.

The transverse Landauer conductance $g^{\mathrm{L}}_{xy}$
can be defined by taking the sample to be a rectangular region 
$[-L^{\ }_{x}/2, +L^{\ }_{x}/2]\times [-L^{\ }_{y}/2, +L^{\ }_{y}/2]$
and attaching four ideal leads to each edge.
As is the case for the Kubo conductivity,
we are going to show that
\begin{equation}
g^{\mathrm{L}}_{xy}=
0.
\label{eq: result for gLxy}
\end{equation}

We now turn to the derivations of Eqs.\
(\ref{eq: result for gLxx}) and (\ref{eq: result for gLxy})
for which we shall set 
\begin{equation}
\hbar =v^{\ }_{F}=-e=1  
\end{equation}
unless 
these constants are written explicitly.

\subsection{
Nazarov formula
           }

Following Refs.%
~\onlinecite{%
Nazarov94,%
Altland05,%
Rejaei96,%
Brouwer96,%
Lamacraft04%
            },
we introduce the generating function for the transmission eigenvalue density
\begin{eqnarray}
{Z}
(\theta^{\ }_{F},\theta^{\ }_{B}):=
\frac{ 
\mathrm{Det}\, 
\left(
1 
- 
\gamma^{\ }_{L}\gamma^{\ }_{R} 
\hat v^{\ }_{L} \mathcal{G}^{R}_{\eta}(0)
\hat v^{\ }_{R} \mathcal{G}^{R}_{\eta}(0)
\right) 
     }
     { 
\mathrm{Det}\, 
\left(
1 
- 
\zeta^{\ }_{L}\zeta^{\ }_{R} 
\hat v^{\ }_{L} \mathcal{G}^{R}_{\eta}(0)
\hat v^{\ }_{R} \mathcal{G}^{R}_{\eta}(0)
\right) }.
\label{eq: def Z(thetaF,thetaB)}
\end{eqnarray}
Here, $\mathrm{Det}$ refers to the functional determinant
over all spatial coordinates
(both inside and outside of the sample)
and spinor indices.
We have also defined 
\begin{eqnarray}
\hat v^{\ }_{R/L} &=& 
{i} \sigma^{\ }_{\pm} \delta(x-x^{\ }_{R/L}).
\end{eqnarray}
Finally, the source terms are parametrized by
$\theta_F$ and $\theta_B$ as
\begin{equation}
\begin{split}
\gamma^{\ }_{R}=
\tan\frac{\theta^{\ }_{F}}{2},
\qquad
\gamma^{\ }_{L}=
\sin\frac{\theta^{\ }_{F}}{2}\cos\frac{\theta^{\ }_{F}}{2},
\\
\zeta^{\ }_{R}=
\tan\frac{{i}\theta^{\ }_{B}}{2},
\qquad
\zeta^{\ }_{L}=
\sin\frac{{i}\theta^{\ }_{B}}{2}\cos\frac{{i}\theta^{\ }_{B}}{2}.
\end{split}
\end{equation}
The Landauer conductance is then given by
\begin{equation}
g^{\mathrm{L}}_{xx}=
\left.
\frac{
\partial Z
     }
     {
\partial
\left(
\gamma^{\ }_{L}
\gamma^{\ }_{R}
\right)
     }
\right|^{\ }_{
\gamma^{\ }_{L}\gamma^{\ }_{R}=
\zeta^{\ }_{L}\zeta^{\ }_{R}=0
             }.
\label{eq: Landauer conductance with Nazarov}
\end{equation}
Furthermore, if the transmission probability
$T^{\ }_{n}$ in channel $n$ (transmission eigenvalue), 
the $n$th positive real-valued eigenvalue of the product of
transmission matrices entering Eq.~(\ref{eq: def Landauer cond})
in descending order,
is written as
\begin{equation}
T^{\ }_{n}=:\cosh^{-2}(\theta^{\ }_{n}/2),
\end{equation}
then the density of transmission eigenvalues,
\begin{subequations}
\label{eq: density of transmission eigenvalues from Nazarov}
\begin{equation}
\rho(\theta):=
\sum_{n}\delta(\theta-\theta^{\ }_{n}),
\label{eq: def rho(theta)}
\end{equation}
is given by
\begin{equation}
\begin{split}
\rho(\theta)=&
\frac{1}{2\pi}
\left[
F(\theta+{i}0^{+}+{i}\pi)
-
F(\theta-{i}0^{+}-{i}\pi)
\right],
\\
F(\theta)=&
\left.
\frac{1}{2}
\left(
\frac{\partial}{\partial \theta^{\ }_{F}}
- 
\frac{\partial}{{i}\partial \theta^{\ }_{B}}
\right)
{Z}(\theta^{\ }_{F},\theta^{\ }_{B})
\right|^{\ }_{\theta^{\ }_{B}=-{i}\theta^{\ }_{F}=\theta}.
\end{split}
\label{eq: density of transmission eigenvalues from Nazarov b}
\end{equation}
\end{subequations}
Once the density of transmission eigenvalues~(\ref{eq: def rho(theta)}) 
is known, we can compute 
the Landauer conductance
(see Ref.~\onlinecite{Beenakker97})
\begin{equation}
g^{\mathrm{L}}_{xx}:=
\sum_{n} T^{\ }_{n},
\end{equation}
the Fano factor
\begin{equation}
F^{\mathrm{L}}_{xx}:=
\frac{
\sum_{n} T^{\ }_{n}\left(1-T^{\ }_{n}\right)
     }
     {
\sum_{n} T^{\ }_{n}
     },
\end{equation}
and other observables in terms of it.

The essential step is to express
the ratio of two determinants (\ref{eq: def Z(thetaF,thetaB)})
by fermionic and bosonic functional integrals as
\begin{subequations}
\label{eq: SUSY rep}
\begin{equation}
\begin{split}
&
{Z}(\theta^{\ }_{F},\theta^{\ }_{B})=
{Z}^{\ }_{F}(\theta^{\ }_{F})
\times
{Z}^{\ }_{B}(\theta^{\ }_{B}),
\\
&
{Z}^{\ }_{F}=
\int\mathcal{D}
\left[
\bar{\chi}^{\ }_{\pm},\chi^{\ }_{\pm}
\right]
e^{-S_F},
\quad
{Z}^{\ }_{B}=
\int\mathcal{D}
\left[
\bar{\varphi}^{\ }_{\pm},\varphi^{\ }_{\pm}
\right]
e^{-S_B},
\end{split}
\end{equation}
where
\begin{equation}
{i}
{S}^{\ }_{F}=
\int_r
\left(
\bar{\chi}^{\ }_{+}, \bar{\chi}^{\ }_{-}
\right)
\left(
\begin{array}{cc}
{i}\eta(r)-\tilde{\mathcal{H}}      &\gamma^{\ }_{R} \hat{v}^{\ }_{R} \\
\gamma^{\ }_{L} \hat{v}^{\ }_{L}    & {i}\eta(r)-\tilde{\mathcal{H}}  \\
\end{array}
\right)
\left(
\begin{array}{c}
{\chi}^{\ }_{+} \\
{\chi}^{\ }_{-}
\end{array}
\right)
\end{equation}
and
\begin{equation}
{i}
{S}^{\ }_{B}=
\int_r
\left(
\bar{\varphi}^{\ }_{+}, \bar{\varphi}^{\ }_{-}
\right)
\left(
\begin{array}{cc}
{i}\eta(r)-\tilde{\mathcal{H}} & \zeta^{\ }_{R}\hat{v}^{\ }_{R}  \\
\zeta^{\ }_{L}\hat{v}^{\ }_{L} & {i}\eta(r)-\tilde{\mathcal{H}}  \\
\end{array}
\right)
\left(
\begin{array}{c}
{\varphi}^{\ }_{+} \\
{\varphi}^{\ }_{-}
\end{array}
\right).
\end{equation}
\end{subequations}
Here,
$\int_r=\int d^2r$ denotes the space integral over the sample and over the leads, 
$(\bar{\chi}^{\ }_{{\alpha}},\chi^{\ }_{{\alpha}})$ 
is a pair of two independent two-component fermionic fields,
and
$(\bar{\varphi}^{\ }_{{\alpha}},\varphi^{\ }_{{\alpha}})$ 
is a pair of two-component
(complex) bosonic fields related by complex conjugation
($\bar{\varphi}_{\alpha}\equiv\varphi_{\alpha}^*$, ${\alpha}=\pm$).
In the functional integral,
$\tilde{\mathcal{H}}$
represents both the sample and leads,
i.e., $\tilde{\mathcal{H}}=\mathcal{H}$ inside the sample.
Similarly, the smearing $\eta(r)$ is zero in the sample but nonvanishing 
in the leads. We now turn to the modeling the leads.

\subsection{
Boundary conditions
           }

There is quite some freedom in modeling the `ideal' leads connected to the
sample. 
In Ref.~\onlinecite{Katsnelson06}
for example,
propagation in the leads is governed by the nonrelativistic
Schr\"odinger equation.
In 
Ref.~\onlinecite{Tworzydlo06}
on the other hand, 
propagation in the leads is governed by the Dirac equation
with a large chemical potential.

We are going to use this freedom 
to choose yet a third model for the leads. 
We demand that a Dirac fermion cannot
exist as a coherently propagating mode in the leads. 
This can be achieved by choosing 
\begin{equation}
\tilde{\mathcal{H}}=
\mathcal{H}
\end{equation}
in both the leads and the sample while using the smearing
to distinguish between the sample and the leads,
\begin{equation}
\eta(r)=
\left\{
\begin{array}{ll}
0,
&
\hbox{ $r$ in the sample},
\\
&
\\
\infty,
&
\hbox{ $r$ in the leads}.
\end{array}
\right.
\label{eq: our model for the leads}
\end{equation}
This choice for modelling
the leads will be justified a posteriori
once we recover from it the results of
Refs.~\onlinecite{Katsnelson06} and \onlinecite{Tworzydlo06}.
(In the following we will use a cylindrical sample.)

The spirit of the choice%
~(\ref{eq: our model for the leads})
is similar to the prescription used
in the nonlinear $\sigma$ model (NL$\sigma$M) 
description of weakly disordered conductors
weakly coupled to ideal leads.%
~\cite{Efetov97} 
In the NL$\sigma$M for the matrix field $Q$, 
leads are represented by a boundary
condition $Q(x= \pm L^{\ }_{x}/2)=\Lambda$ where
$\Lambda$ is a fixed matrix in the symmetric space
of which $Q$ is an element.
The matrix $Q$ describes the interacting diffusive modes 
of a weakly disordered metal.
In a loose sense one may be able to think of this boundary condition
as prohibiting coherent propagation of these diffusive modes in the leads.

For a metallic sample (with a finite Fermi surface) in the 
ballistic regime that is weakly coupled to the leads, 
charge transport is strongly dependent on the nature of the contacts 
and the leads. On the other hand, for a metallic sample 
in the diffusive regime and not too large couplings to the leads, 
the conductance is mostly determined by the disordered region itself.
(See Ref.\ \onlinecite{Nikolic01} and references therein.) 
The conductivity of ballistic Dirac fermions in two dimensions is
of order 1. Transport should thus behave in a way similar 
to that in a diffusive metal.\cite{Tworzydlo06}
We would then expect that the microscopic modeling of the leads
should have little effects on the conductance; i.e., the
conductance should depend only on the intrinsic properties of the
two-dimensional sample such as the conductivity. Reassuringly, 
it has been observed by Schomerus that transport in graphene is 
largely independent of the microscopic modeling of the leads.%
\cite{Schomerus06}
Correspondingly, we will show that our model for the coupling between
the sample and the reservoirs
(\ref{eq: our model for the leads}) leads to conformal invariant 
(i.e., scale-invariant and hence a renormalization group fixed point) 
boundary conditions to the supersymmetric field theory 
(\ref{eq: SUSY rep}).

The condition%
~(\ref{eq: our model for the leads})
suggests that the effects of the leads 
are equivalent to singling out 
a special configuration of the fields in the leads
through the condition of a saddle point.
To investigate the saddle-point
condition implied by the leads%
~(\ref{eq: our model for the leads}),
we introduce first the chiral base 
$\psi^{\dag}$, $\psi$, $\beta^{\dag}$, and $\beta$ defined by
\begin{equation}
\begin{split}
&
\left(
\psi^{\dag}_{{\alpha}},
\bar{\psi}^{\dag}_{{\alpha}}
\right)=
\bar{\chi}^{\ }_{{\alpha}}\sigma^{\ }_{x}
\sqrt{2\pi},
\qquad
\left(
\begin{array}{c}
\psi^{\ }_{{\alpha}} \\
\bar{\psi}^{\ }_{{\alpha}}
\end{array}
\right)=
\sqrt{2\pi}\,
\chi^{\ }_{{\alpha}},
\\
&
\left(
\beta^{\dag}_{{\alpha}},
\bar{\beta}^{\dag}_{{\alpha}}
\right)=
\bar{\varphi}^{\ }_{{\alpha}}\sigma^{\ }_{x}
\sqrt{2\pi},
\qquad
\left(
\begin{array}{c}
\beta^{\ }_{{\alpha}} \\
\bar{\beta}^{\ }_{{\alpha}}
\end{array}
\right)=
\sqrt{2\pi}\,
\varphi^{\ }_{{\alpha}},
\end{split}
\end{equation}
with ${\alpha}=\pm$ in terms of which
\begin{subequations}
\label{eq: final path integral rep of Nazarov generating Z}
\begin{equation}
\begin{split}
&
{Z}^{\ }_{F}=
\int\mathcal{D}
\left[
\bar{\psi}^{\ }_{\pm},\psi^{\ }_{\pm}
\right]
e^{-S_F^{(0)}-S_F^{\eta}-S_F^{\gamma}},
\quad
\\
&
{Z}^{\ }_{B}=
\int\mathcal{D}
\left[
\bar{\beta}^{\ }_{\pm},\beta^{\ }_{\pm}
\right]
e^{-S_B^{(0)}-S_B^{\eta}-S_B^{\zeta}},
\end{split}
\end{equation}
where
\begin{eqnarray}
S_{F}^{(0)}
&=&
\sum_{{\alpha}=\pm}
\int_r \frac{1}{\pi}
\left(
\psi^{\dag}_{{\alpha}}
\bar{\partial}
\psi^{\ }_{{\alpha}}
+
\bar{\psi}^{\dag}_{{\alpha}}
{\partial}
\bar{\psi}^{\ }_{{\alpha}}
\right),
\nonumber\\
S_F^{\eta}&=&
\sum_{{\alpha}=\pm}
\int_r
\frac{\eta(r)}{2\pi}
\left(
\psi^{\dag}_{{\alpha}}\bar{\psi}^{\ }_{{\alpha}}
+
\bar{\psi}^{\dag}_{{\alpha}}\psi^{\ }_{{\alpha}}
\right),
\\
S_F^{\gamma}&=&
\int_r
\left[
\frac{\gamma^{\ }_{R}}{\pi}
\bar{\psi}^{\dag}_{+}\bar{\psi}^{\ }_{-}
\delta(x-x^{\ }_{R})
+
\frac{\gamma^{\ }_{L}}{\pi}
\psi^{\dag}_{-}\psi^{\ }_{+}
\delta(x-x^{\ }_{L})
\right]
\nonumber
\end{eqnarray}
and
\begin{eqnarray}
S^{(0)}_{B}&=&
\sum_{{\alpha} = \pm}
\int_r \frac{1}{\pi}
\left(
\beta^{\dag}_{{\alpha}}
\bar{\partial}
\beta^{\ }_{{\alpha}}
+
\bar{\beta}^{\dag}_{{\alpha}}
{\partial}
\bar{\beta}^{\ }_{{\alpha}}
\right),
\nonumber\\
S_{B}^{\eta}
&=&
\sum_{{\alpha}=\pm}
\int_r
\frac{\eta(r)}{2\pi}
\left(
\beta^{\dag}_{{\alpha}}\bar{\beta}^{\ }_{{\alpha}}
+
\bar{\beta}^{\dag}_{{\alpha}}\beta^{\ }_{{\alpha}}
\right),
 \\
S_{B}^{\zeta}
&=&
\int_r
 \left[
\frac{\zeta^{\ }_{R}}{\pi}
\bar{\beta}^{\dag}_{+}\bar{\beta}^{\ }_{-}
\delta(x-x^{\ }_{R})
+
\frac{\zeta^{\ }_{L}}{\pi}
\beta^{\dag}_{-}\beta^{\ }_{+}
\delta(x-x^{\ }_{L})
\right].
\nonumber 
\end{eqnarray}
\end{subequations}
The actions ${S}^{(0)}_{F}$ 
and ${S}^{(0)}_{B}$ give 
two copies of the Dirac fermion CFT ($c=1$) 
and bosonic ghost CFT ($c=-1$), respectively.%
~\cite{Friedan86,Guruswamy00}

In the leads, the field entering the functional integrals must then satisfy
\begin{equation}
\sum_{{\alpha}=\pm}
\left(
\psi^{\dag}_{{\alpha}}\bar{\psi}^{\ }_{{\alpha}}
+
\bar{\psi}^{\dag}_{{\alpha}}\psi^{\ }_{{\alpha}}
+
\beta^{\dag}_{{\alpha}}\bar{\beta}^{\ }_{{\alpha}}
+
\bar{\beta}^{\dag}_{{\alpha}}\beta^{\ }_{{\alpha}}
\right)
=0.
\label{eq: saddle point eq implied by our def of leads}
\end{equation}
Possible solutions to the
saddle-point equations~(\ref{eq: saddle point eq implied by our def of leads})
are
\begin{equation}
\psi^{\ }_{{\alpha}} = \pm {i} \bar{\psi}^{\ }_{{\alpha}},
\qquad
\beta^{\ }_{{\alpha}} = \pm {i} \bar{\beta}^{\ }_{{\alpha}},
\qquad
{\alpha}=\pm.
\label{eq: solutions to SP}
\end{equation}
Not all solutions~(\ref{eq: solutions to SP})
yield the desired Landauer conductance. 
One choice that does,
as will be shown in 
Secs.\ \ref{subsec: Landauer conductance} 
and \ref{subsec: twisted partition functions}
below,
 amounts to the boundary conditions
\begin{equation}
\label{eq: BC due to leads on psi's and beta's}
\begin{split}
&
\psi^{\ }_{\alpha}(x= \mp L^{\ }_{x}/2,y)
=
\mp {i}
\bar{\psi}^{\ }_{\alpha}(x= \mp L^{\ }_{x}/2,y),
\\
&
\beta^{\ }_{\alpha} (x= \mp L^{\ }_{x}/2,y)
=
\mp {i}
\bar{\beta}^{\ }_{\alpha} (x= \mp L^{\ }_{x}/2,y),
\end{split}
\end{equation}
with $\alpha=\pm$.
These boundary conditions break the factorization into
a holomorphic and antiholomorphic sector present in the bulk.
This is not to say that conformal invariance is broken, however,
as it is possible to eliminate one sector  
(say the antiholomorphic one)
altogether
in favor of the other
(say holomorphic),
thereby yielding a chiral conformal field theory.%
\cite{%
Cardy84%
     }

At last, we need to impose antiperiodic boundary conditions
in the (periodic) $y$ direction of the cylinder,
\begin{equation}
\label{eq: BC due cylinder}
\begin{split}
&
\left(
\begin{array}{c}
\psi^{\ }_{{\alpha}} 
\\
\bar{\psi}^{\ }_{{\alpha}} 
\\
\end{array}
\right)(x,y)
=
-
\left(
\begin{array}{c}
\psi^{\ }_{{\alpha}} 
\\
\bar{\psi}^{\ }_{{\alpha}} 
\end{array}
\right)(x,y+L^{\ }_{y}),
\\
&
\left(
\begin{array}{c}
\beta^{\ }_{{\alpha}} 
\\
\bar{\beta}^{\ }_{{\alpha}} 
\end{array}
\right)(x,y)
=
-
\left(
\begin{array}{c}
\beta^{\ }_{{\alpha}} 
\\
\bar{\beta}^{\ }_{{\alpha}} 
\end{array}
\right)(x,y+L^{\ }_{y}),
\end{split}
\end{equation}
for ${\alpha}=\pm$ and $-L^{\ }_{x}/2<x<+L^{\ }_{x}/2$ and 
$0\leq y<L^{\ }_{y}$.
Our choice of antiperiodic boundary conditions
for the fermionic fields $\psi^{\ }_{\alpha}$ and $\bar{\psi}^{\ }_{\alpha}$
is the natural one if the $y$ direction is 
thought of as representing a ``time'' coordinate.
The choice of periodic boundary conditions can be implemented 
at the price of introducing an additional operator in the 
conformal field theory. However, the Einstein conductivity does
not depend on this choice of boundary conditions.

\subsection{
Landauer conductance 
           }
\label{subsec: Landauer conductance}

Before using the generating function%
~(\ref{eq: def Z(thetaF,thetaB)})
to compute directly the density of transmission eigenvalues%
~(\ref{eq: density of transmission eigenvalues from Nazarov}),
we compute the Landauer conductance%
~(\ref{eq: Landauer conductance with Nazarov})
as a warm-up. Insertion of Eq.%
~(\ref{eq: final path integral rep of Nazarov generating Z})
into Eq.%
~(\ref{eq: Landauer conductance with Nazarov})
yields
\begin{equation}
\label{eq: Landauer xx as 4-point fct in free susy}
g^{\mathrm{L}}_{xx}=
\frac{1}{\pi^2}
\oint\limits_{L^{\ }_{y}}^{0}
\oint\limits_0^{L^{\ }_{y}}
dy dy'
\left\langle
\bigl (
\psi^{\   }_{+} 
\psi^{\dag}_{-}
\bigr )
\left(r\right)
\bigl (
\bar{\psi}^{\   }_{-} 
\bar{\psi}^{\dag}_{+}
\bigr )
\left(r'\right)
\right\rangle^{\ }_{0}
\end{equation}
with 
$r =(-L^{\ }_{x}/2,y)$ 
and 
$r'=(+L^{\ }_{x}/2,y')$.
The expectation value
$
\left\langle
\cdots
\right\rangle^{\ }_{0}
$
is performed here with the action
$S_{F}^{(0)}+S_{B}^{(0)}$
from 
Eqs.\ (\ref{eq: final path integral rep of Nazarov generating Z})
supplemented with the boundary conditions
(\ref{eq: BC due to leads on psi's and beta's})
and
(\ref{eq: BC due cylinder}).
The four-fermion correlation function in Eq.%
~(\ref{eq: Landauer xx as 4-point fct in free susy})
can be expressed in terms of
two-point correlation functions given by
\begin{subequations}
\label{eq: generic 2-pt fct in free susy}
\begin{equation}
\begin{split}
\left\langle 
\psi^{\   }_{{\alpha}}(x,y)
\psi^{\dag}_{{\alpha}'}(0,0) 
\right\rangle^{\ }_{0}=&
\left\langle 
\beta^{\   }_{{\alpha}}(x,y)
\beta^{\dag}_{{\alpha}'}(0,0) 
\right\rangle^{\ }_{0}
\\
=&
\,\delta^{\ }_{{\alpha},{\alpha}'}
\mathcal{G}^{\ }_{0}(x,y;L^{\ }_{x},L^{\ }_{y})
\end{split}
\end{equation}
with ${\alpha},{\alpha}'=\pm$ and where%
~\cite{footnote: method image}
\begin{equation}
\mathcal{G}^{\ }_{0}(x,y;L^{\ }_{x},L^{\ }_{y}):=
\sum_{m\in\mathbb{Z}}
\frac{
(-1)^m 
     }
     {
\frac{L^{\ }_{y}}{\pi} 
\sinh 
\frac{\pi}{L^{\ }_{y}}
\left(
x
+
{i}y
+
2mL^{\ }_{x}
\right)
     }.
\end{equation}
\end{subequations}
After combining Eq.%
~(\ref{eq: Landauer xx as 4-point fct in free susy})
with Eq.%
~(\ref{eq: generic 2-pt fct in free susy}),
one finds
\begin{equation}
g^{\mathrm{L}}_{xx}=
2    
\sum_{n=0}^{\infty}
\cosh^{-2}
\left(
(2n+1)\frac{\pi L^{\ }_{x}}{L^{\ }_{y}}
\right).
\label{eq: result for Landauer conductance}
\end{equation}
Each transmission eigenvalue
\begin{equation}
T^{\ }_{n}:=
\cosh^{-2} 
\left(
(2n+1)\frac{\pi L^{\ }_{x}}{L^{\ }_{y}}
\right),
\qquad
n=0,1,2,\cdots,
\end{equation}
is twofold degenerate.
We shall see that this degeneracy originates 
from the two species ($\alpha=\pm$)
when deducing 
\begin{equation}
\theta^{\ }_{n}=
(2n+1)\frac{2\pi L^{\ }_{x}}{L^{\ }_{y}},
\qquad
n=0,1,2,\cdots,
\end{equation}
directly from Eq.%
~(\ref{eq: density of transmission eigenvalues from Nazarov}).

The Landauer conductance~(\ref{eq: result for Landauer conductance})
is a monotonic decreasing function of $L^{\ }_{x}/L^{\ }_{y}$.
When the sample is the surface of a long and narrow cylinder,
$L^{\ }_{x}/L^{\ }_{y}\gg1$, the conductance is dominated by the contribution
from the smallest transmission eigenvalue and decays exponentially fast
with $L^{\ }_{x}/L^{\ }_{y}\gg1$,
\begin{equation}
g^{\mathrm{L}}_{xx}=
2    
e^{-2\pi L^{\ }_{x}/L^{\ }_{y}}
+
\mathcal{O}
\left(
e^{-6\pi L^{\ }_{x}/L^{\ }_{y}}
\right).
\end{equation}
In the opposite limit of a very short cylinder,
$L^{\ }_{x}/L^{\ }_{y}\ll1$,
\begin{equation}
\begin{split}
g^{\mathrm{L}}_{xx}=&    
2\frac{L^{\ }_{y}}{2\pi L^{\ }_{x}}
\int\limits_0^{\infty} 
\frac{d\lambda}{\cosh^2\lambda}
+
\mathcal{O}
\left[
\left(
L^{\ }_{y}/L^{\ }_{x}
\right)^{0}
\right]
\\
=&\,
\frac{1}{\pi}
\frac{L^{\ }_{y}}{ L^{\ }_{x}}
+
\mathcal{O}
\left[
\left(
L^{\ }_{y}/L^{\ }_{x}
\right)^{0}
\right].
\end{split}
\end{equation}

We now turn to the computation of $g^{\mathrm{L}}_{xy}$.
To this end, we take the sample to be a rectangular region 
$[-L^{\ }_{x}/2, +L^{\ }_{x}/2]\times [-L^{\ }_{y}/2, +L^{\ }_{y}/2]$
and attach ideal leads to each edge.
Instead of the antiperiodic boundary condition
(\ref{eq: BC due cylinder}),
we must treat the boundary conditions along the $y$ direction on equal footing
with the boundary conditions along the $x$ direction; i.e., we impose the 
boundary conditions
\begin{equation}
\label{eq: BC due to leads on psi's and beta's in y direction}
\begin{split}
&
\psi^{\ }_{\alpha} (x,y=\mp L^{\ }_{y}/2)
=
\mp {i}
\bar{\psi}^{\ }_{\alpha} (x,y=\mp L^{\ }_{y}/2),
\\
&
\beta^{\ }_{\alpha} (x,y=\mp L^{\ }_{y}/2)
=
\mp {i}
\bar{\beta}^{\ }_{\alpha} (x,y=\mp L^{\ }_{y}/2),
\end{split}
\end{equation}
with $\alpha=\pm$ together with 
the boundary conditions%
~(\ref{eq: BC due to leads on psi's and beta's}).
Equations 
(\ref{eq: bilocal conductivity at omega 0 in terms Greens functions})
and (\ref{eq: def Landauer cond}),
when applied to $g^{\mathrm{L}}_{xy}$,
give
\begin{equation}
\begin{split}
g^{\mathrm{L}}_{xy}=&
\frac{{i}}{2\pi^2}
\int\limits_{-L^{\ }_{y}/2}^{+L^{\ }_{y}/2} dy
\int\limits_{-L^{\ }_{x}/2}^{+L^{\ }_{x}/2} dx'
\Big\langle
\left[
\psi^{\dag}_{+}
\psi^{\ }_{-}
(r')
+
\bar{\psi}^{\dag}_{+}
\bar{\psi}^{\ }_{-}
(r')
\right]
\\
&
\times
\left[
\psi^{\dag}_{-}
\psi^{\ }_{+}
(r)
-
\bar{\psi}^{\dag}_{-}
\bar{\psi}^{\ }_{+}
(r)
\right]
\Big\rangle_0,
\end{split}
\end{equation}
where
$r =(-L^{\ }_{x}/2,y)$ 
and 
$r'=(x',+L^{\ }_{y}/2)$.
The expectation value
$
\left\langle
\cdots
\right\rangle^{\ }_{0}
$
is performed here with the action
$S^{(0)}_{F}+S^{(0)}_{B}$
supplemented with the boundary conditions%
~(\ref{eq: BC due to leads on psi's and beta's})
and
~(\ref{eq: BC due to leads on psi's and beta's in y direction}).
Using these boundary conditions, 
we can remove the right movers at the interfaces 
$x=-L^{\ }_{x}/2$ and $y=+L^{\ }_{y}/2$
with the result 
\begin{equation}
g^{\mathrm{L}}_{xy}=0.
\end{equation}

\subsection{
Twisted partition functions
           }
\label{subsec: twisted partition functions}

We now go back to the direct calculation of the density
of transmission eigenvalues,
Eq. (\ref{eq: def rho(theta)}),
from Eq.%
~(\ref{eq: density of transmission eigenvalues from Nazarov b}).
We proceed in two steps.

First, we perform a gauge transformation
(defined in Appendix 
~\ref{appsec: Unitary transformation that decouples the pm grading})
on the integration variables in the fermionic and bosonic path integrals,
respectively, that diagonalizes the fermionic and bosonic actions
\begin{equation}
\begin{split}
&
{S}^{(0)}_{F}
+
S_F^{\gamma}
\to
{S}^{+}_{F}
+
{S}^{-}_{F},
\\
&
{S}^{(0)}_{B}
+
S_{B}^{\zeta}
\to
{S}^{+}_{B}
+
{S}^{-}_{B}.
\end{split}
\end{equation}
In doing so the boundary conditions%
~(\ref{eq: BC due to leads on psi's and beta's}) 
that implement the presence of the leads are changed to
\begin{subequations}
\label{eq: new BC due to leads on psi's and beta's}
\begin{equation}
\begin{split}
&
\psi^{\ }_{\alpha} (x=-L^{\ }_{x}/2,y)
=
-{i}
\bar{\psi}^{\ }_{\alpha} (x=-L^{\ }_{x}/2,y),
\\
&
\psi^{\ }_{\alpha} (x=+L^{\ }_{x}/2,y)
=
+{i}
e^{+{i}\alpha \theta^{\ }_{F}}
\bar{\psi}^{\ }_{\alpha} (x=+L^{\ }_{x}/2,y)
\end{split}
\end{equation}
and
\begin{equation}
\begin{split}
&
\beta^{\ }_{\alpha} (x=-L^{\ }_{x}/2,y)
=
-{i}
\bar{\beta}^{\ }_{\alpha} (x=-L^{\ }_{x}/2,y),
\\
&
\beta^{\ }_{\alpha}(x=+L^{\ }_{x}/2,y)
=
+{i}
e^{+\alpha \theta^{\ }_{B}}
\bar{\beta}^{\ }_{\alpha} (x=+L^{\ }_{x}/2,y),
\end{split}
\end{equation}
\end{subequations}
with $\alpha=\pm$,
for the ``gauge-transformed'' fields.

Second, we introduce the four independent
partition functions
${Z}^{+}_{F}$,
${Z}^{-}_{F}$,
${Z}^{+}_{B}$,
and
${Z}^{-}_{B}$
describing two species ($\pm$) of fermionic ($F$)
and bosonic ($B$) free fields
(with holomorphic and antiholomorphic components)
satisfying the boundary conditions%
~(\ref{eq: new BC due to leads on psi's and beta's})
and%
~(\ref{eq: BC due cylinder}).
These are equivalent to four independent partition functions
anti-${Z}^{+}_{F;\mathrm{ch}}$,
${Z}^{-}_{F;\mathrm{ch}}$,
${Z}^{+}_{B;\mathrm{ch}}$,
and
${Z}^{-}_{B;\mathrm{ch}}$
describing free holomorphic fields that fulfill the
boundary conditions
\begin{subequations}
\label{eq: final BC's}
\begin{equation}
\begin{split}
\psi^{\ }_{\pm}(x,y+L^{\ }_{y})=&
-
\psi^{\ }_{\pm}(x,y),
\\
\psi^{\ }_{\pm}(x+2 L^{\ }_{x},y)=&
-
e^{\pm{i}\theta^{\ }_{F}}\psi^{\ }_{\pm}(x,y),
\end{split}
\end{equation}
and
\begin{equation}
\begin{split}
\beta^{\ }_{\pm}(x,y+L^{\ }_{y})=&
-
\beta^{\ }_{\pm}(x,y),
\\
\beta^{\ }_{\pm}(x+2L^{\ }_{x},y)=&
-
e^{\pm \theta^{\ }_{B}}\beta_{\pm}^{\ }(x,y),
\end{split}
\end{equation}
\end{subequations}
with $-L^{\ }_{x}\leq x<L^{\ }_{x}$ and $0\leq y<L^{\ }_{y}$.%
\cite{Cardy84}
We have thus traded the antiholomorphic sector
in favor of a cylinder twice as long and a change in the boundary
conditions%
~(\ref{eq: new BC due to leads on psi's and beta's})
implementing the presence of the leads.

According to Ref.%
~\onlinecite{Guruswamy96},
the chiral partition functions
${Z}^{\pm}_{F;\mathrm{ch}}$
and
${Z}^{\pm}_{B;\mathrm{ch}}$
are given by
\begin{eqnarray}
&&
{Z}^{\pm}_{F;\mathrm{ch}}=
q^{\frac{-1}{24}}
\prod_{n=0}^{\infty}
\big(
1+ e^{\pm {i}\theta^{\ }_{F}}
q^{n+\frac{1}{2}}
\big)
\big(
1+ e^{\mp {i}\theta^{\ }_{F}}
q^{n+\frac{1}{2}}
\big),
\nonumber\\
&&\\
&&
{Z}^{\pm}_{B;\mathrm{ch}}=
q^{\frac{+1}{24}}
\prod_{n=0}^{\infty}
\frac{1}{ 1+ e^{\pm \theta^{\ }_B} 
q^{n+\frac{1}{2}}}
\frac{1}{ 1+ e^{\mp \theta^{\ }_B} 
q^{n+\frac{1}{2}}},
\nonumber
\end{eqnarray}
up to factors that cancel each other 
when we combine the fermionic and bosonic partition functions.
We have introduced the variable
$
q:=e^{-2\pi\times2L^{\ }_{x}/L^{\ }_{y}}.
$
We have
${Z}^{{\alpha}}_{{F}}{Z}^{{\alpha}}_{{B}}=1$,
separately for each species ${\alpha}=\pm$,
when the boundary conditions are the same for fermions and bosons, 
i.e., when ${i}\theta^{\ }_{F}=\theta^{\ }_{B}$,
as it should be a consequence of global supersymmetry.
With the help of
\begin{eqnarray}
&&
{Z}^{\pm}_{F;\mathrm{ch}}\!\!=\!\!
\prod_{n=0}^{\infty}
\left(
1
+ 
q^{\frac{2n+1}{2}}
\right)^{2}
\prod_{n=0}^{\infty}
\!
\left[
1
- 
\frac{\sin^{2} 
\frac{
\theta^{\ }_{F}}{2} 
     }
     {
\cosh^{2} \frac{\pi (2n+1) L^{\ }_{x}}{L^{\ }_{y}} 
     }
\right],
\nonumber\\&&\\&&
{Z}^{\pm}_{B;\mathrm{ch}}\!\!=
\!\!\prod_{n=0}^{\infty}\!\!
\left(
1+ 
q^{\frac{2n+1}{2}}
\right)^{\!-2}\!
\prod_{n=0}^{\infty}\!
\left[
1- 
\frac{\sin^{2} 
\frac{
{i}\theta^{\ }_{B}}{2} 
     }
     {
\cosh^{2} \frac{\pi (2n+1) L^{\ }_{x}}{L^{\ }_{y}} 
     }
\right]^{-1}
\!\!\!\!\!\!,
\nonumber
\end{eqnarray}
one verifies that
\begin{eqnarray}
\rho(\theta)&=&
2
\sum_{n=0}^{\infty}
\delta
\left(
\theta
-
(2n+1)
\frac{2 \pi L^{\ }_{x}}{L^{\ }_{y}}
\right).
\end{eqnarray}
The origin of the twofold degeneracy is the fact 
that the $\pm$ species have decoupled.

As it should be, the Landauer conductance is
\begin{equation}
\begin{split}
g^{L}_{xx}&=
\int d\theta\, 
\rho(\theta) 
\cosh^{-2} 
\theta/2
\\
&=
2\sum_{n=0}^{\infty}
\cosh^{-2}
\left(
(2n+1)\frac{\pi L^{\ }_{x}}{L^{\ }_{y}}
\right).
\end{split}
\label{eq: reproducing the Landauer conductance}
\end{equation}

Equation~(\ref{eq: reproducing the Landauer conductance})
implies that the density of transmission eigenvalues
is uniform,
\begin{equation}
\rho(\theta)=
\frac{L^{\ }_{y}}{\pi L^{\ }_{x}},
\end{equation}
in the limit ${\pi L^{\ }_{x}}\ll{L^{\ }_{y}}$.
In this sense, the transmission eigenvalue density 
for the massless Dirac equation in a sample with 
the topology of a short cylinder agrees with that
of a disordered metallic wire in the diffusive regime.%
~\cite{Beenakker97}
This is why transport for ballistic Dirac fermions 
is similar to mesoscopic transport in disordered quantum wires.

The generating function technique can also be applied to
$g^{\mathrm{L}}_{xy}$.
If we follow the discussions for
 $g^{\mathrm{L}}_{xx}$,
one simply finds that the partition
function is actually independent of $\theta^{\ }_{F,B}$.

\section{
Chiral disorder with time-reversal symmetry
         }

We devote this section to calculating the first-order correction
to the Einstein conductivity induced by
a weak real-valued random (white-noise) hopping amplitude 
between nearest-neighbor sites of the honeycomb lattice
at the band center. We are going to show that the 
Einstein conductivity is unchanged to this order.
We then go on to show that the Einstein conductivity is an 
analytic function of the disorder strength.

We start from a single spinless fermion hopping 
between nearest-neighbor sites of the honeycomb lattice at the band center.
The hopping amplitudes are assumed real
with small random fluctuations compared to their uniform mean. 
This model was introduced by Foster and Ludwig in Ref.~\onlinecite{Foster06}.

For weak disorder, this model can be simplified by linearizing the spectrum
of the clean limit at the band center. In this approximation the clean spectrum
is that of two flavors of Dirac fermions, each Dirac fermion encoding 
the low-energy and long-wavelength description of the conduction band 
in the valley with a Fermi point.
Weak disorder induces both intravalley and intervalley scatterings
whose effects involve the two flavors of Dirac fermions.
The characteristic disorder strength for intranode scattering
is $g^{\ }_{\mathrm{A}}$; that for internode scattering is
$g^{\ }_{\mathrm{M}}$.\cite{Foster06,Hatsugai97} 
Without loss of generality, 
we shall set $g^{\ }_{\mathrm{A}}=0$ and concentrate on
$g^{\ }_{\mathrm{M}}>0$.%
\cite{footnote: gauging away random vector potential}
In a fixed realization of the disorder, 
the single-particle Green's functions at the band center 
can be derived as correlation functions
for fermionic 
$(\psi^{\dag},\psi)$
and bosonic (ghost)
$(\beta^{\dag},\beta)$
variables
from the partition function
$Z=Z^{\ }_{F}\times Z^{\ }_{B}$ with
\begin{subequations}
\label{eq: final partition fct HWK}
\begin{equation}
\begin{split}
&
Z^{\ }_{F}[m,\bar m,A,\bar A]= 
  \int\mathcal{D}[\psi^{\dag},\psi]\
  e^{
    -S^{\ }_{F}-S^{\eta}_{F} 
    },
\\
&
Z^{\ }_B[m,\bar m,A,\bar A] =
  \int\mathcal{D}[\beta^{\dag},\beta]\
  e^{
    -S^{\ }_{B}-S^{\eta}_{B}
    }.
\end{split}
\label{eq: final partition fct HWK a}
\end{equation}
The low-energy effective (Dirac) action for the fermionic part is given by
\begin{equation}
\begin{split}
    S^{\ }_{F}
    =&\,
    \int_r\frac{1}{2\pi}
    \sum_{a=1}^{2}
    \Big[
    \psi^{a\dag}
    (2\bar{\partial}+\bar{A})
    \psi^{ \   }_{a}
    +
    \bar\psi^{a \dag}
    (2\partial+A)
    \bar\psi^{   \  }_{a}
\\
&
\qquad
\qquad
    +
    \bar{m}\psi^{a\dag}\bar\psi^{\ }_{a}
    +    
         m \bar\psi^{a\dag}\psi^{\ }_{a}
    \Big]
\end{split}
\end{equation}
and
\begin{equation}
\begin{split}
    S^{\eta}_{F}
    =&
\int_r
\frac{{i}\eta}{2\pi}
    \left(
      \psi^{1\dag}\bar\psi^{2\dag}
      +
      \bar\psi^{1\dag}\psi^{2\dag}
      -
      \psi^{\ }_{2}\bar\psi^{\ }_{1}
      -
      \bar\psi^{\ }_{2}\psi^{\ }_{1}
    \right).
\end{split}
\end{equation}
The low-energy effective action for the bosonic part is given by
the replacement $S^{\ }_{F}\to S^{\ }_{B}$ under
\begin{equation}
\left(
\psi^{a\dag},
\bar
\psi^{a\dag},
\psi^{\ }_{a},
\bar
\psi^{\ }_{a}
\right)
\longrightarrow
\left(
\beta^{a\dag},
\bar
\beta^{a\dag},
\beta^{\ }_{a},
\bar
\beta^{\ }_{a}
\right)
\end{equation}
with $a=1,2$ and
  \begin{equation}
    S^{\eta}_{B}=
   \int_r
\frac{  {i}\eta }{2\pi}
\left(
      -
      \beta^{1\dag}\bar\beta^{2\dag}
      -
      \bar\beta^{1\dag}\beta^{2\dag}
      -
      \beta^{\ }_{2}\bar\beta^{\ }_{1}
      -
      \bar\beta^{\ }_{2}\beta^{\ }_{1}
\right).
\end{equation}
The Abelian gauge fields $A$ and $\bar{A}$ are source terms 
for the paramagnetic response function.
The disorder is realized by the complex-valued random mass
$m$ and its complex conjugate $\bar{m}$ which obey the
distribution law
\begin{eqnarray}
P[m]\propto
\int\mathcal{D}[\bar{m},m]\,
\exp
\left(
-
\frac{1}{2 g^{\ }_{\mathrm{M}}}
\int_r
\bar{m}m
\right).
\label{eq: prob distribution m}
\end{eqnarray}
The time-reversal symmetry of the lattice model has become
\begin{equation}
\begin{split}
&
\psi^{1\dag} \to -\psi^{2\dag},
\,\,
\psi^{2\dag} \to +\psi^{1\dag},
\,\,
\psi_{1}^{\ } \to -\psi_{2}^{\ },
\,\,
\psi_{2}^{\ } \to +\psi_{1}^{\ },
\\
&
\beta^{1\dag} \to -{i}\beta^{2\dag},
\,\,
\beta^{2\dag} \to +{i}\beta^{1\dag},
\,\,
\beta_{1}^{\ } \to +{i}\beta_{2}^{\ },
\,\,
\beta_{2}^{\ } \to -{i}\beta_{1}^{\ },
\end{split}
\end{equation}
with the same transformation laws for the fields 
with overbars.
The sublattice symmetry
of the lattice model has become
\begin{equation}
\begin{split}
&
\psi^{a\dag}\to
\psi^{a \dag},
\,\,
\bar{\psi}^{a\dag}\to
\bar{\psi}^{a\dag},
\,\,
\psi^{\ }_{a}\to
-\psi^{\ }_{a},
\,\,
\bar{\psi}^{\ }_{a}\to
-\bar{\psi}^{\ }_{a},
\\
&
\beta^{a\dag}\to
\beta^{a\dag},
\,\,
\bar{\beta}^{a\dag}\to
\bar{\beta}^{a\dag},
\,\,
\beta^{\ }_{a}\to
-\beta^{\ }_{a},
\,\,
\bar{\beta}^{\ }_{a}\to
-\bar{\beta}^{\ }_{a},
\end{split}
\end{equation}
\end{subequations}
for $a=1,2$.

We want to compute the first-order correction
to the mean Einstein conductivity.
To this end, we must double the number of integration variables
in Eq.~(\ref{eq: final partition fct HWK a}).
This is so because the response function
is the product of two single-particle Green's functions.
This is achieved by extending the range
$a=1,2$ of the flavor index in Eq.~(\ref{eq: final partition fct HWK})
to $\mathfrak{a}:=(a\,\alpha)$ with $\alpha=\pm$ being a color index, 
one for each of the two single-particle Green's functions.
[The same doubling of integration variables was introduced
in Eq.~(\ref{eq: SUSY rep}).] We shall also use the more compact
notation by which the capital latin index
``$\mathrm{A}$'' replaces the original two-flavor indices $a=1,2$ 
in that it also carries a grade which is either 0 
when we want to refer to bosons -- 
say, $\chi^{\ }_{(\mathrm{A}\alpha)}=\beta^{\ }_{(a\alpha)}$ --
or 1 when we want to refer to fermions --
say, $\chi^{\ }_{(\mathrm{A}\alpha)}=\psi^{\ }_{(a\alpha)}$.
It is the grade of the indices $\mathrm{A}$ that enters 
expressions such as $(-)^{\mathrm{A}}$.
Correspondingly, we shall use the collective index
$\mathfrak{A}:=(\mathrm{A}\alpha)$
to treat bosons and  fermions with the flavor
index $a=1,2$ and the color index $\alpha=\pm$ at once.

Since we are only after the mean response function
(and not higher moments), it can be obtained
from 
\begin{equation}
\begin{split}
&
Z[A,\bar A]:=
\int\mathcal{D}
\left[
\chi^{\dag},
\chi,
\bar{\chi}^\dag,
\bar{\chi}
\right]
\exp
\left(
-
S^{\ }_{*}
-
S^{\ }_{\eta}
-
S^{\ }_{\mathrm{M}}
\right),
\\
&
S^{\ }_{*}:=
\int_{r}\frac{1}{2\pi}
\left[
\chi^{\mathfrak{A}\dag}
\left(2\bar\partial+\bar A\right)
\chi^{\ }_{\mathfrak{A}}
+
\bar \chi^{\mathfrak{A}\dag}
\left(2\partial+A\right)
\bar \chi^{\ }_{\mathfrak{A}}
\right],
\\
&
S^{\ }_{\eta}:=
\int_{r}
\frac{i\eta}{2\pi}
\mathcal{O}^{\ }_{\eta},
\\
&
S^{\ }_{\mathrm{M}}:=
\int_{r}
\frac{g^{\ }_{\mathrm{M}}}{2\pi^{2}}
\mathcal{O}^{\ }_{\mathrm{M}}.
\end{split}
\label{eq: action S* Seta SM}
\end{equation}
Here, the interaction induced by integrating over the
random mass with the probability distribution%
~(\ref{eq: prob distribution m})
is described by
\begin{equation}
\begin{split}
\mathcal{O}^{\ }_{\mathrm{M}}:=
\chi^{\ }_{\mathfrak{A}}
\chi^{\mathfrak{B}\dag}
\bar
\chi^{\ }_{\mathfrak{B}}
\bar
\chi^{\mathfrak{A}\dag}
(-1)^{\mathrm{A}},
\end{split}
\end{equation}
while the smearing bilinear is
\begin{equation}
\begin{split}
\mathcal{O}^{\ }_{\eta}:=&
\delta^{\ }_{1,\mathrm{A}}
\delta^{\ }_{1,\mathrm{B}}
\left[
\chi^{\mathfrak{A}\dag}
\left(i\tau^{\ }_{y}\right)^{\ }_{\mathrm{AB}}
\bar
\chi^{\mathfrak{B}\dag}
+
\chi^{\ }_{\mathfrak{A}}
\left(i\tau^{\ }_{y}\right)^{\mathrm{AB}}
\bar
\chi^{\ }_{\mathfrak{B}}
\right]
\\
&
-
\delta^{\ }_{0,\mathrm{A}}
\delta^{\ }_{0,\mathrm{B}}
\left[
\chi^{\mathfrak{A}\dag}
\left(\tau^{\ }_{x}\right)^{\ }_{\mathrm{AB}}
\bar
\chi^{\mathfrak{B}\dag}
\!+\!
\chi^{\ }_{\mathfrak{A}}
\left(\tau^{\ }_{x}\right)^{\mathrm{AB}}
\bar
\chi^{\ }_{\mathfrak{B}}
\right].
\end{split}
\end{equation}

According to Eqs.%
~(\ref{eq: basis independent representation Einstein conductivity})
and
(\ref{eq: jmu is a commutator}),
the Einstein conductivity~(\ref{eq: equivalence with Einstein cond at T=0})
can be expressed in terms of the current-current correlation function
$
-(\pi^{2}\hbar^{2}/e^{2})\Sigma^{\ }_{xx}(r,0)
$. The latter function can be chosen to be represented in terms 
of bosonic variables, yielding
\begin{subequations}
\begin{equation}
\begin{split}
&
\Big\langle
\left(
J_{(1-)}^{\prime (1+)}
+
\bar{J}_{(1-)}^{\prime (1+)} 
\right)(r)
\left(
J_{(2-)}^{\prime (2+)}
+
\bar{J}_{(2-)}^{\prime (2+)} 
\right)(0)
\\
&\,
\hphantom{\frac{e^{2}}{\pi^{2}\hbar^{2}}}
+
\left(
J_{(2+)}^{\prime (2-)}
+
\bar{J}_{(2+)}^{\prime (2-)} 
\right)(r)
\left(
J_{(1+)}^{\prime (1-)}
+
\bar{J}_{(1+)}^{\prime (1-)} 
\right)(0)
\Big\rangle.
\end{split}
\end{equation}
[This is a generalization of Eq.~(3.6) 
in which the single-particle Green's functions and currents 
are $4\times4$ matrices.] 
Here, we have introduced the currents
\begin{equation}
\begin{split}
J_{(a\alpha)}^{\prime (a'\alpha')}:=&
\beta^{\ }_{(a\alpha)}
\,
\beta^{(a'\alpha')\dag},
\qquad
\bar J_{(a\alpha)}^{\prime (a'\alpha')}:=
\bar \beta^{\ }_{(a\alpha)}
\,
\bar \beta^{(a'\alpha')\dag},
\end{split}
\label{eq: def currents needed for transport}
\end{equation}
with the flavor indices
$a,a'=1,2$
and the color indices
$\alpha,\alpha'=\pm$,
while the expectation value refers to 
\begin{equation}
\langle
\left(
\cdots
\right)
\rangle
:=
\int\mathcal{D}
\left[
\chi^{\dag},
\chi,
\bar{\chi}^\dag,
\bar{\chi}
\right]
\exp
\left(
-
S^{\ }_{0}
-
S^{\ }_{\mathrm{M}}
\right)
\left(
\cdots
\right)
\end{equation}
\end{subequations}
with
$
S^{\ }_{0}:=
S^{\ }_{*}
+
S^{\ }_{\eta}.
$
In the clean limit, the Einstein conductivity 
\begin{equation}
\mathrm{Re}\,\sigma^{\ }_{xx}:=
\frac{\hbar}{4\pi L^{2}}
\int_{r}
\int_{r'}
\Sigma^{\ }_{xx}(r,r')
\label{eq: Einstein for HWK}
\end{equation}
is given by twice the value of
Eq.~(\ref{eq: order of limit III for dc Kubo})
at zero temperature. This is understood as follows.
The bilocal conductivity reduces to
computing the free-field expectation values with the action $S^{\ }_{0}$ 
of bilinears in the normal-ordered current%
~(\ref{eq: def currents needed for transport}).
By Wick's theorem, the bilocal conductivity can be reduced 
to the products of pairs of free-field propagators. 
The relevant free-field propagators are
\begin{widetext}
\begin{subequations}
\begin{equation}
\begin{split}
&
\langle
\beta^{(2\alpha)\dag}(r)\bar\beta^{(1\alpha')\dag}(0)
\rangle=
\langle
\bar\beta^{(2\alpha)\dag}(r)\beta^{(1\alpha')\dag}(0)
\rangle=
\langle
\beta_{(1\alpha)}^{\ }(r)\bar\beta_{(2\alpha')}^{\ }(0)
\rangle=
\langle
\bar\beta_{(1\alpha)}^{\ }(r)\beta_{(2\alpha')}^{\ }(0)
\rangle=
\delta^{\ }_{\alpha\alpha'}
2\pi 
\int_k
e^{{i}r\cdot k}
\frac{
{i}\eta 
     }
     {
\eta^2 
+ 
k^2
     }
\end{split}
\label{eq: two point fcts a}
\end{equation}
and
\begin{equation}
\begin{split}
&
\langle
\beta^{(a\alpha)\dag}(r)\beta_{(a'\alpha')}^{\ }(0)
\rangle=
\delta^{\ }_{aa'}\,
\delta^{\ }_{\alpha\alpha'}\,
2\pi{i}
\int_k
e^{{i}r\cdot k}
\frac{
k^{\ }_{x} 
-
{i}k^{\ }_{y} 
     }
     {
\eta^2 
+ 
k^2
     },
\qquad
\langle
\bar\beta^{(a\alpha)\dag}(r)\bar\beta_{(a'\alpha')}^{\ }(0)
\rangle=
\delta^{\ }_{aa'}\,
\delta^{\ }_{\alpha\alpha'}\,
2\pi{i}
\int_k
e^{{i}r\cdot k}
\frac{
k^{\ }_{x} 
+
{i}k^{\ }_{y} 
     }
     {
\eta^2 
+ 
k^2
     },
\end{split}
\label{eq: two point fcts b}
\end{equation}
\end{subequations}
\end{widetext}
with $a,a'=1,2$ and $\alpha,\alpha'=\pm$ for the Einstein conductivity.

We turn next to the first-order correction in powers
of $g^{\ }_{\mathrm{M}}$ of the mean Einstein conductivity
and show that it vanishes. This is understood as follows.
By translation invariance, the integration over $r'$ 
in Eq.~(\ref{eq: Einstein for HWK})
yields
\begin{subequations}
\label{eq: doing the r' integration}
\begin{equation}
\mathrm{Re}\,\sigma^{\ }_{xx}=
\frac{\hbar}{4\pi}
\int_{r}
\Sigma^{\ }_{xx}(r,0),
\end{equation}
where the bilocal conductivity can be decomposed into three
contributions,
\begin{equation}
\Sigma^{\ }_{xx}(r,0)=
\Sigma^{(\mathrm{hh})}_{xx}(r,0)
+
\Sigma^{(\mathrm{aa})}_{xx}(r,0)
+
\Sigma^{(\mathrm{ha})}_{xx}(r,0).
\end{equation}
\end{subequations}
The holomorphic contribution is
\begin{subequations}
\label{eq: def Sigma hh}
\begin{equation}
\begin{split}
\Sigma^{(\mathrm{hh})}_{xx}(r,0)=&\,
\frac{-e^{2}}{\pi^{2}\hbar^{2}}
B^{\ \mathfrak{a}_2^{\ },\ \mathfrak{a}_4^{\ }}
 _{\mathfrak{a}_1^{\ } ,\ \mathfrak{a}_3^{\ }}
\left\langle
J^{\prime \mathfrak{a}_1^{\ }}_{\mathfrak{a}_2^{\ }}(r)
\,
J^{\prime \mathfrak{a}_{3}^{\ }}_{\mathfrak{a}_4^{\ }}(0)
\right\rangle
\end{split}
\end{equation}
[summation convention over repeated indices 
$\mathfrak{a}=(a\alpha)$
is assumed on the right-hand side]
with
\begin{equation}
B^{\ (1-),\ (2-)}
 _{  (1+),\ (2+)}
=
B^{\ (2+),\ (1+)}
 _{  (2-),\ (1-)}
=
1
\end{equation}
\end{subequations}
the only nonvanishing coefficients.
The antiholomorphic contribution is
\begin{subequations}
\label{eq: def Sigma aa}
\begin{equation}
\begin{split}
\Sigma^{(\mathrm{aa})}_{xx}(r,0)=&\,
\frac{-e^{2}}{\pi^{2}\hbar^{2}}
\bar
B^{\ \mathfrak{a}_2^{\ },\ \mathfrak{a}_4^{\ }}
 _{\mathfrak{a}_1^{\ },\ \mathfrak{a}_3^{\ }}
\left\langle
\bar J^{\prime \mathfrak{a}_1^{\ }}_{\mathfrak{a}_2^{\ }}(r)
\,
\bar J^{\prime \mathfrak{a}_3^{\ }}_{\mathfrak{a}_4^{\ }}(0)
\right\rangle,
\end{split}
\end{equation}
with
\begin{equation}
\bar
B^{\ (1-),\ (2-)}
 _{  (1+),\ (2+)}
=
\bar
B^{\ (2+),\ (1+)}
 _{  (2-),\ (1-)}
=
1
\end{equation}
\end{subequations}
the only nonvanishing coefficients.
The mixed holomorphic and antiholomorphic contribution is
\begin{subequations}
\label{eq: def Sigma ha}
\begin{equation}
\begin{split}
\Sigma^{(\mathrm{ha})}_{xx}(r,0)=&\,
\frac{-e^{2}}{\pi^{2}\hbar^{2}}
C^{\ \mathfrak{a}_2^{\ },\ \mathfrak{a}_4^{\ }}
 _{ \mathfrak{a}_1^{\ },\ \mathfrak{a}_3^{\ }}
\left\langle
J^{\prime \mathfrak{a}_1^{\ }}_{\mathfrak{a}_2^{\ }}(r)
\,
\bar 
J^{\prime \mathfrak{a}_3^{\ }}_{\mathfrak{a}_4^{\ }}(0)
\right\rangle,
\end{split}
\end{equation}
with
\begin{equation}
\begin{split}
1=&
C^{\ (1-),\ (2-)}
 _{  (1+),\ (2+)}
=
C^{\  (2-),\ (1-)}
 _{ (2+),\ (1+)}
\\
=&
C^{\ (2+),\ (1+)}
 _{  (2-),\ (1-)}
=
C^{\ (1+),\ (2+)}
 _{  (1-),\ (2-)}
\end{split}
\end{equation}
\end{subequations}
the only nonvanishing coefficients.
The two-point functions%
~(\ref{eq: def Sigma hh})
and~(\ref{eq: def Sigma aa})
transform irreducibly and nontrivially under a rotation
of the Euclidean plane $r\in\mathbb{R}^{2}$.
Consequently, their separate contributions to%
Eq.~(\ref{eq: doing the r' integration})
are vanishing and
\begin{equation}
\mathrm{Re}\,\sigma^{\ }_{xx}=
\frac{\hbar}{4\pi}
\int_{r}
\Sigma^{(\mathrm{ha})}_{xx}(r,0).
\end{equation}

The first-order correction to the Einstein conductivity
in the clean limit is
\begin{equation}
\begin{split}
\delta\,\mathrm{Re}\,\sigma^{\ }_{xx}:=&\,
+
\frac{g^{\ }_{\mathrm{M}}}{2\pi^{2}}
\frac{e^{2}}{2\pi^{2}h}
C^{\ \mathfrak{a}_2^{\ },\ \mathfrak{a}_4^{\ }}
 _{  \mathfrak{a}_1^{\ },\ \mathfrak{a}_3^{\ }}
\\
&
\times
\int_{r}
\int_{r'}
\left\langle
J^{\prime \mathfrak{a}_1^{\ }}_{\mathfrak{a}_2^{\ }}(r)
\,
\bar 
J^{\prime \mathfrak{a}_3^{\ }}_{\mathfrak{a}_4^{\ }}(0)
\,
\mathcal{O}^{\ }_{\mathrm{M}}(r')
\right\rangle^{\ }_{0}.
\end{split}
\label{eq: first order correction}
\end{equation}
Summation convention over repeated indices is assumed on the right-hand side.
Carrying the double integration in
Eq.~(\ref{eq: first order correction})
yields, with the help of Wick's theorem, 
\begin{equation}
\begin{split}
\delta\,\mathrm{Re}\,\sigma^{\ }_{xx}=&\,
\frac{g^{\ }_{\mathrm{M}}}{2\pi^{2}}
\frac{e^{2}}{2\pi^{2}h}
C^{\ \mathfrak{a}_2^{\ },\ \mathfrak{a}_1^{\ }}
 _{  \mathfrak{a}_1^{\ },\ \mathfrak{a}_2^{\ }}
\times
\pi^{2}.
\label{eq: first order correction bis}
\end{split}
\end{equation}
Summation convention over repeated indices is assumed on the right-hand side.
Since
$
C^{\ \mathfrak{a}_2^{\ },\ \mathfrak{a}_1^{\ }}
 _{  \mathfrak{a}_1^{\ },\ \mathfrak{a}_2^{\ }}
=0
$
for any pair 
$
(a^{\ }_{1}\alpha^{\ }_{1})
$
and
$
(a^{\ }_{2}\alpha^{\ }_{2})
$,
it follows that
\begin{equation}
\delta\,\mathrm{Re}\,\sigma^{\ }_{xx}=0.
\label{eq: first order correction final}
\end{equation}

Observe here that the factor
$\pi^{2}$
in Eq.~(\ref{eq: first order correction bis})
comes from the spatial integrations.
It follows that the first-order correction to the Einstein
conductivity is free from a logarithmic dependence on the
ultraviolet cutoff. 
The correction of order $g^{2 }_{\mathrm{M}}$
is also free from a logarithmic divergence 
but nonvanishing.\cite{Ostrovsky06}
These results are special cases of the fact
that the Einstein conductivity \textit{must be}
an analytic function of the coupling constant $g^{\ }_{\mathrm{M}}$.
Indeed, it was shown in Ref.~\onlinecite{Guruswamy00}
that the action $S^{\ }_{*}+S^{\ }_{\mathrm{M}}$
of Eq.~(\ref{eq: action S* Seta SM})
has the symmetry group GL($4|4$) while the sector  
of the theory that carries no U(1) Abelian charge,
the so-called PSL($4|4$) sector, is a critical theory.
One consequence of this is that
the beta function for $g^{\ }_{\mathrm{M}}$ 
vanishes to all orders in  
$g^{\ }_{\mathrm{M}}$
as $\mathcal{O}^{\ }_{\mathrm{M}}$
belongs to the PSL($4|4$) sector.
Another consequence is that the Einstein conductivity must be
an analytic function of $g^{\ }_{\mathrm{M}}$ as
the bilocal conductivity also belongs to the PSL($4|4$) sector.%
\cite{footnote: belongs to PSL}

\section{
Conclusions
        }

We have shown how to compute the transmission eigenvalues 
for a single massless Dirac fermion propagating freely in two dimensions
within a two-dimensional conformal field theory description
in the presence of 
twisted boundary conditions. We hope that this derivation, which
is complementary to the ones from 
Refs.~\onlinecite{Katsnelson06} and \onlinecite{Tworzydlo06}
using direct methods of quantum mechanics,
can be generalized to the presence of certain types of disorder
so as to obtain nonperturbative results.

We have also shown that the Einstein conductivity,
which is obtained from the regularization of the 
dc Kubo conductivity in terms of the four possible products
of advanced and retarded Green's functions by taking the
dc limit \textit{before} removing the smearing
in the single-particle Green's functions,
agrees with the conductivity determined from the Landauer formula.

Finally, we noted that, as a consequence of the fixed point theory
discussed in Ref.~\onlinecite{Guruswamy00}, the Einstein conductivity
is an analytic function of the strength $g^{\ }_{\mathrm{M}}$
of the disorder which preserves the sublattice symmetry of 
the random hopping model on the honeycomb lattice.
Moreover, the first-order correction in $g^{\ }_{\mathrm{M}}$
to the Einstein conductivity was shown to vanish.

\section*{Acknowledgments}

We would like to thank A.\ D.\ Mirlin for useful discussions.
S.R. would like to thank Kazutaka Takahashi for useful discussions.
C.M. and A.F. acknowledge hospitality of the Kavli Institute for 
Theoretical Physics at Santa Barbara during the completion
of the manuscript.
This research was supported in part by the National Science Foundation
under Grant No.\ PHY99-07949 and by a Grant-In-Aid for Scientific Research
(No.\ 16GS0219) from MEXT of Japan.

\appendix

\section{
Numerical integration of Eq.~(\ref{eq: bilocal if single particle H 2})
when $\mu=\nu=x$
        }
\label{app sec: preparation for numerical integration}

\begin{widetext}
To evaluate numerically Eq.~(\ref{eq: bilocal if single particle H 2})
when $\mu=\nu=x$ and for any finite $\eta>0$
it is useful to perform the integration over momenta in
\begin{equation}
\begin{split}
\mathrm{Re}\,\sigma^{\ }_{xx}(\omega,\eta,\beta)=&
\frac{(ev^{\ }_{F})^2}{(2\pi)^2\omega}
\int\limits^{+\infty}_{-\infty}d\varepsilon\,
[f^{\ }_\beta(\varepsilon+\hbar\omega)-f^{\ }_\beta(\varepsilon)]
\int\limits^{+\infty}_{0} 
dk\,k\,
\left(
\frac{\varepsilon-i\hbar\eta}
     {(\varepsilon-i\hbar\eta)^2-(\hbar v_Fk)^2}
-
\frac{\varepsilon+i\hbar\eta}
     {(\varepsilon+i\hbar\eta)^2-(\hbar v_Fk)^2}
\right)
\\
&
\,\times
\left(
\frac{\varepsilon+\hbar\omega-i\hbar\eta}
     {(\varepsilon+\hbar\omega-i\hbar\eta)^2-(\hbar v_Fk)^2}
-
\frac{\varepsilon+\hbar\omega+i\hbar\eta}
     {(\varepsilon+\hbar\omega+i\hbar\eta)^2-(\hbar v_Fk)^2}
\right).
\end{split}
\end{equation}
This gives
\begin{subequations}
\label{eq: integrating over momenta}
\begin{equation}
\begin{split}
\frac{h}{e^2}
\mathrm{Re}\,\sigma^{\ }_{xx}(X,Y)=&
\frac{1}{2\pi}\frac{1}{X}
\int\limits^{+\infty}_{-\infty}da\,
\left[
f^{\ }_{\beta=1}(q/Y)
-
f^{\ }_{\beta=1}(p/Y)
\right]
\left[
\left(
\frac{X}{4a(a^2+1)}
+
\frac{4a}{X(X^2+4)}
\right)
\frac{1}{2}\ln\frac{1+p^2}{1+q^2}
\right.
\\
&\,
\left.
+
\left(
\frac{X}{4(a^2+1)}
+
\frac{1}{X}
\right)
\left(
\mathrm{arctan}\,p
-
\mathrm{arctan}\,q
\right)
-
\left(
\frac{2a}{X^2+4}
+
\frac{1}{2a}
\right)
\left(
\mathrm{arctan}\,p
+
\mathrm{arctan}\,q
\right)
\right],
\end{split}
\label{eq: integrating over momenta a}
\end{equation}
where
\begin{equation}
X:=\frac{\omega}{\eta},
\qquad
Y:=\frac{1}{\beta\hbar\eta},
\qquad
a:=\frac{\varepsilon}{\hbar\eta},
\qquad
p:= 
a-\frac{X}{2},
\qquad
q:= 
a+\frac{X}{2}.
\label{eq: integrating over momenta b}
\end{equation}
\end{subequations}

\end{widetext}

\section{
Gauge transformation 
        }
\label{appsec: Unitary transformation that decouples the pm grading}

The ``gauge field'' $\gamma^{\ }_{R/L}$
can be removed by a suitable gauge transformation,
$\psi \to  U_F^{\ }(x)\psi$,
$
\psi^{\dag} \to  \psi^{\dag}U_{F}^{\ }(x)^{-1},
$
where
$
U_F^{\ } (x)
=
V_{F}^{\ }
W_{F}^{\ }(x)$.
The gauge transformation $W^{\ }_F(x)$ is position dependent,
off-diagonal in the $\pm$ sector,
and given by
\begin{equation}
\begin{split}
A_F(x)=&
\left(
\begin{array}{cc}
0 
& 
\gamma^{\ }_{R}\sigma^{\ }_{d}\delta(x-x^{\ }_{R}) 
\\
\gamma^{\ }_{L}\sigma^{\ }_{u}\delta(x-x^{\ }_{L}) 
& 
0
\end{array}
\right)^{\ }_{\pm},
\\
W_F^{\ }(x)
=&
T^{\ }_{x}
\exp
\left(
+
\int_{x^{\ }_{L}}^{x} dx' 
A_F^{\ }(x')
\right),
\end{split}
\end{equation}
where $T^{\ }_{x}$ represents $x$ ordering,
and
\begin{eqnarray}
\sigma^{\ }_{d} &=&
\left(
\begin{array}{cc}
0 &  0 \\
0 & 2
\end{array}
\right)^{\ }_{R/L},
\quad
\sigma^{\ }_{u} =
\left(
\begin{array}{cc}
2 & 0 \\
0 & 0
\end{array}
\right)^{\ }_{R/L}.
\end{eqnarray}
Here, $\cdots^{\ }_{R/L}$ denotes the holomorphic/antiholomorphic sector
whereas $\cdots^{\ }_{\pm}$ represents the $\pm$ sector.
On the other hand, $V_F^{\ }$ is position independent,
diagonal in  the holomorphic and antiholomorphic sectors, and given by
\begin{equation}
V_F^{\ }=
\left(
\begin{array}{cc}
a_{F}^{\ } & 0 \\
0 & a_{F}^{\ }
\end{array}
\right)^{\ }_{R/L},
a_F^{\ } =
\left(
\begin{array}{cc}
e^{+{i}\theta_F}-1 & e^{-{i}\theta_F}-1 \\
\sin\theta_F & \sin \theta_F
\end{array}
\right)^{\ }_{\pm}.
\end{equation}
Since the action in the bulk is diagonal in the $\pm$ sector,
$V$ does not affect the action in the bulk, while it is chosen to
diagonalize the boundary conditions at both ends of the cylinder.

The ``gauge field'' $\zeta^{\ }_{R/L}$ can be removed 
with the help of the gauge transformation,
$\beta \to  V_{B}^{\ }W_{B}^{\ }(x)\beta$,
$
\beta^{\dag} \to  \beta^{\dag}
W_{B}^{\ }(x)^{-1}V_{B}^{-1}
$
where
\begin{equation}
\begin{split}
A_B(x)=&
\left(
\begin{array}{cc}
0 
& 
\zeta^{\ }_{R}\sigma^{\ }_{d}\delta(x-x^{\ }_{R}) 
\\
\zeta^{\ }_{L}\sigma^{\ }_{u}\delta(x-x^{\ }_{L}) 
& 
0
\end{array}
\right)^{\ }_{\pm},
\\
W_B^{\ }(x)
=&
T^{\ }_{x}
\exp
\left(
+
\int_{x^{\ }_{L}}^{x} dx' 
A_B^{\ }(x')
\right)
\end{split}
\end{equation}
and
\begin{equation}
V_B^{\ }=
\left(
\begin{array}{cc}
a_{B}^{\ } & 0 \\
0 & a_{B}^{\ }
\end{array}
\right)^{\ }_{R/L},
\
a_B^{\ } =
\left(
\begin{array}{cc}
e^{+\theta_B}-1 & e^{-\theta_B}-1 \\
{i}\sin\theta_B & {i}\sin \theta_B
\end{array}
\right)^{\ }_{\pm}.
\end{equation}


\begin{thebibliography}{99}

\bibitem{Novoselov04}
K.\ S.\ Novoselov,
A.\ K.\ Geim,
S.\ V.\ Morozov,
D.\ Jiang,
Y.\ Zhang,
S.\ V.\ Dubonos,
I.\ V.\ Grigorieva, and
A.\ A.\ Firsov,
Science \textbf{306}, 666 (2004).

\bibitem{Novoselov05}
K.\ S.\ Novoselov,
A.\ K.\ Geim,
S.\ V.\ Morozov,
D.\ Jiang,
M.\ I.\ Katsnelson,
I.\ V.\ Grigorieva, 
S.\ V.\ Dubonos, and
A.\ A.\ Firsov,
Nature (London) \textbf{438}, 197 (2005).

\bibitem{Zhang05a}
Y.\ Zhang,
J.\ P.\ Small,
M.\ E.\ S.\ Amori, and
P.\ Kim,
Phys.\ Rev.\ Lett.\ \textbf{94}, 176803 (2005).

\bibitem{Zhang05b}
Y.\ Zhang,
V.\ -W.\ Tan,
H.\ L.\ Stormer, and
P.\ Kim,
Nature (London) \textbf{438}, 201 (2005).

\bibitem{Katsnelson06} 
M.\ I.\ Katsnelson,
Eur.\ Phys.\ J.\ B \textbf{51}, 157 (2006).

\bibitem{Tworzydlo06}
J.\ Tworzydlo, B.\ Trauzettel,  M.\ Titov, A.\ Rycerz, 
and C.\ W.\ Beenakker,
Phys.\ Rev.\ Lett.\ \textbf{96}, 246802 (2006).

\bibitem{Edwards71} 
J.\ T.\ Edwards and D.\ J.\ Thouless, 
J.\ Phys.\ C \textbf{4}, 453 (1971).

\bibitem{Nazarov94} 
Yu.\ V.\ Nazarov, 
Phys.\ Rev.\ Lett.\ \textbf{73}, 134 (1994).

\bibitem{Rejaei96}
B.\ Rejaei,
Phys.\ Rev.\ B \textbf{53}, R13235 (1996).

\bibitem{Brouwer96}
P.\ W.\ Brouwer and K.\ Frahm,
Phys.\ Rev.\ B \textbf{53}, 1490 (1996).

\bibitem{Lamacraft04}
A.\ Lamacraft, B.\ D.\ Simons, and M.\ R.\ Zirnbauer,
Phys.\ Rev.\ B \textbf{70}, 075412 (2004).

\bibitem{Altland05}
A.\ Altland, A.\ Kamenev, and C.\ Tian,
Phys.\ Rev.\ Lett.\ \textbf{95}, 206601 (2005).

\bibitem{unitary Q1D wire}
M.\ R.\ Zirnbauer, 
Phys.\ Rev.\ Lett.\ \textbf{69}, 1584 (1992);
A.\ D.\ Mirlin, A.\ M\"uller-Groeling, and M.\ R.\ Zirnbauer, 
Ann.\ Phys.\ (N.Y.) \textbf{236}, 325 (1994);
K.\ Frahm, Phys.\ Rev.\ Lett.\ \textbf{74}, 4706 (1995).

\bibitem{Mudry99}
C.\ Mudry, P.\ W.\ Brouwer, and A.\ Furusaki, 
Phys.\ Rev.\ B \textbf{59}, 13221 (1999).

\bibitem{Brouwer00}
P.\ W.\ Brouwer, A.\ Furusaki, I.\ A.\ Gruzberg, and C.\ Mudry, 
Phys.\ Rev.\ Lett.\ \textbf{85}, 1064 (2000).

\bibitem{Ludwig94} 
A.\ W.\ W.\ Ludwig, M.\ P.\ A.\ Fisher, R.\ Shankar, and G.\ Grinstein, 
Phys.\ Rev.\ B \textbf{50}, 7526 (1994).

\bibitem{Cserti06}
J.\ Cserti,
Phys.\ Rev.\ B \textbf{75}, 033405 (2007).

\bibitem{Ziegler07}
K.\ Ziegler, 
\texttt{cond-mat/0701300} (unpublished).

\bibitem{Fradkin86}
E.\ Fradkin, 
Phys.\ Rev.\ B \textbf{33}, 3263 (1986).

\bibitem{Lee93}
P.\ A.\ Lee, Phys.\ Rev.\ Lett.\ \textbf{71}, 1887 (1993).

\bibitem{Shon98}
N.\ H.\ Shon and T.\ Ando,
J.\ Phys.\ Soc.\ Jpn.\ \textbf{67}, 2421 (1998).

\bibitem{Durst99} 
A.\ C.\ Durst and P.\ A.\ Lee, 
Phys.\ Rev.\ B \textbf{62}, 1270 (2000).

\bibitem{Gorbar02}
E.\ V.\ Gorbar,
V.\ P.\ Gusynin,
V.\ A.\ Miransky,
and
I.\ A.\ Shovkovy,
Phys.\ Rev.\ B \textbf{66}, 045108 (2002).

\bibitem{Peres05}
N.\ M.\ R.\ Peres,
F.\ Guinea,
and 
A.\ H.\  Castro Neto,
Phys.\ Rev.\ B \textbf{73}, 125411 (2006).

\bibitem{Guruswamy00}
S.\ Guruswamy, A.\ LeClair, and A.\ W.\ W.\ Ludwig,
Nucl.\ Phys.\  B \textbf{583}, 475 (2000).

\bibitem{Ostrovsky06}
P.\ M.\ Ostrovsky, I.\ V.\ Gornyi, and A.\ D.\ Mirlin,
Phys.\ Rev.\ B\ \textbf{74}, 235443 (2006)

\bibitem{Verbaarschot94}
J. J. M. Verbaarschot, 
Phys.\ Rev.\ Lett.\ \textbf{72}, 2531 (1994).

\bibitem{Zirnbauer96}   
M.\ R.\ Zirnbauer,
J.\ Math.\ Phys.\ \textbf{37}, 4986 (1996). 

\bibitem{Altland97}   
A. Altland and M. R. Zirnbauer, 
Phys.\ Rev.\ B \textbf{55}, 1142 (1997).

\bibitem{Heinzner05}
P. Heinzner, A. Huckleberry, and M. R. Zirnbauer,
Commun.\ Math.\ Phys.\ \textbf{257}, 725 (2005).

\bibitem{Aleiner06}
I. L. Aleiner and K. B. Efetov,
Phys.\ Rev.\ Lett.\ \textbf{97}, 236801 (2006).

\bibitem{Altland06}
Alexander Altland, 
Phys.\ Rev.\ Lett.\ \textbf{97}, 236802 (2006).

\bibitem{Ando02}
T.\ Ando and H.\ Suzuura,
J.\ Phys.\ Soc.\ Jpn.\ \textbf{71},
2753 (2002);
H.\ Suzuura and T.\ Ando,
Phys.\ Rev.\ Lett.\ \textbf{89}, 266603 (2002).

\bibitem{Nomura07}
Kentaro Nomura and A. H. MacDonald,
Phys.\ Rev.\ Lett.\ \textbf{98}, 076602 (2007).

\bibitem{Rycerz07}   
A. Rycerz, J. Tworzydlo, and C. W. J. Beenakker,
cond-mat/0612446 (unpublished).

\bibitem{Mahan}
G.\ D.\ Mahan,
\textit{Many-particle physics}, 3rd ed.,
(Kluwer Academic, New York, 2000).

\bibitem{Baranger89}
H.\ U.\ Baranger and A.\ D.\ Stone,
Phys.\ Rev.\ B \textbf{40}, 8169 (1989).

\bibitem{footnote: a dot b}
We are using the notation
$
a\cdot b\equiv a^{\ }_{\mu}b^{\ }_{\mu}
$
between any two vectors $a,b\in\mathbb{R}^{2}$.
Integration over space (momentum)
is sometimes abreviated by
$\int_{r}\equiv\int d^2r$
$\left(\int_{k}\equiv\int\frac{d^2 k}{(2\pi)^2}\right)$.

\bibitem{footnote: role of zero modes}
Observe that only single-particle eigenstates
(\ref{eq: single-particle eigenstates})
with a nonvanishing energy eigenvalue contribute to
Eq.\ (\ref{eq: ac Kubo conductivity b}). 
In particular,
eigenstates with a vanishing energy eigenvalue, i.e., zero modes, 
do not contribute to Eq.~(\ref{eq: ac Kubo conductivity b})
for $\omega>0$. This should be contrasted with Eqs.%
~(\ref{eq: order of limit III for dc Kubo})
and~(\ref{eq: result for Landauer conductance}) 
for which zero modes contribute.
The localization properties of zero modes for
lattices with sublattice symmetry 
and their relation to the conductance at the band center
after connecting the lattices to leads have been studied by
P.\ W.\ Brouwer, E.\ Racine, A.\ Furusaki, Y.\ Hatsugai, Y.\ Morita, 
and C.\ Mudry, Phys.\ Rev.\ B \textbf{66}, 014204 (2002).

\bibitem{Kramer93}
B.\ Kramer and A.\ MacKinnon,
Rep.\ Prog.\ Phys.\ \textbf{56}, 1469 (1993).

\bibitem{McKane81}
A.\ J.\ McKane and M.\ Stone, 
Ann.\ Phys.\ (N. Y.) \textbf{131}, 36 (1981).

\bibitem{footnote: Durst}
To compare our result with the one by Durst and Lee, 
note that it is the conserved spin current in a $d$-wave superconductor
that plays, in Ref.~\onlinecite{Durst99}, the role of our charge current.

\bibitem{OtherNumericsFigOne}
See also Figs.\ 7 and 8 of Ref.\ \onlinecite{Peres05},
where the (momentum and frequency-dependent) imaginary part of
the self-energy, which corresponds to our smearing $\eta$,
is determined in a certain way self-consistently in the presence of 
disorder and the electron-electron interaction.

\bibitem{Wallace47}
P.\ R.\ Wallace,
Phys.\ Rev.\ \textbf{71}, 622 (1947).

\bibitem{footnote: ideal leads}
The leads that we shall consider in this paper are ideal
in the sense that they describe a perfect conductor,
i.e., one with diverging (infinite) dc conductivity:
for example, a nearest-neighbor tight-binding Hamiltonian 
on a bipartite lattice at the band center 
known to possess chiral (Refs.\ \onlinecite{Verbaarschot94,Zirnbauer96,Altland97,Heinzner05}) symmetry.

\bibitem{Fisher81}
D.\ S.\ Fisher and P.\ A.\ Lee,
Phys.\ Rev.\ B \textbf{23}, 6851 (1981).

\bibitem{Xiong96} 
See S.\ Xiong, N.\ Read, and A.\ D.\ Stone, 
Phys.\ Rev.\ B \textbf{56}, 3982 (1996) and references therein.

\bibitem{Beenakker97}
C.\ W.\ J.\ Beenakker,
Rev.\ Mod.\ Phys.\ \textbf{69}, 731 (1997).

\bibitem{Efetov97} 
K.\ Efetov, 
\textit{Supersymmetry in disorder and chaos}, 
(Cambridge University Press, New York, 1997).

\bibitem{Nikolic01}
B.\ K.\ Nikoli\'c,
Phys.\ Rev.\  B \textbf{64}, 165303 (2001).

\bibitem{Schomerus06}
H.\ Schomerus,
\texttt{cond-mat/0611209} (unpublished).

\bibitem{Friedan86}
D.\ Friedan, E.\ Martinec, and S.\ Shenker,
Nucl.\ Phys.\ B \textbf{271}, 93 (1986).

\bibitem{Cardy84}
J.\ L.\ Cardy, 
Nucl.\ Phys.\ B \textbf{240}, 514 (1984);
I.\ Affleck, 
\textit{ibid.} B \textbf{336}, 517 (1990);
I.\ Affleck and A.\ W.\ W.\ Ludwig, 
\textit{ibid.} B \textbf{360}, 641 (1991).

\bibitem{footnote: method image}
The method of images can be used here
to solve the Dirac equation in the complex plane.

\bibitem{Guruswamy96}
S.\ Guruswamy and A.\ W.\ W.\ Ludwig,
Nucl.\ Phys.\ B \textbf{519}, 661 (1998).

\bibitem{Foster06}
M.\ S.\ Foster and A.\ W.\ W.\ Ludwig,
Phys.\ Rev.\ B \textbf{73}, 155104 (2006).

\bibitem{Hatsugai97}
The same effective low-energy description is obtained
for spinless fermions hopping on a lattice with $\pi$ flux
per plaquette as shown by
Y.\ Hatsugai, X.-G.\ Wen, and M.\ Kohmoto, 
Phys.\ Rev.\ B \textbf{56}, 1061 (1997).

\bibitem{footnote: gauging away random vector potential}
A random imaginary vector potential is also generated in the continuum
approximation by the nearest-neighbor real-valued random hopping.
This source of disorder can here be ``gauged away'' using manipulations
similar to the ones introduced in Ref.~\onlinecite{Guruswamy96}.

\bibitem{footnote: belongs to PSL}
As shown in Ref.~\onlinecite{Guruswamy96},
it is sufficient to notice
that the bilocal conductivity depends on the off-diagonal components
of the GL($4|4$) currents 
(see, e.g., Table 2 of Ref.~\onlinecite{Mudry03})
to establish that the bilocal conductivity belongs to the PSL($4|4$) sector.

\bibitem{Mudry03}
C.\ Mudry, S.\ Ryu, and A.\ Furusaki,
Phys.\ Rev.\ B \textbf{67}, 064202 (2003).

\end{thebibliography}
\end{document}